%% file: main.tex
\documentclass{article}

\usepackage[nonatbib, preprint]{neurips_2025}

\usepackage[utf8]{inputenc} % allow utf-8 input
\usepackage[T1]{fontenc}    % use 8-bit T1 fonts
\usepackage{url}            % simple URL typesetting
\usepackage{booktabs}       % professional-quality tables
\usepackage{amsfonts}       % blackboard math symbols
\usepackage{nicefrac}       % compact symbols for 1/2, etc.
\usepackage{microtype}      % microtypography
\usepackage{xcolor}         % colors
\usepackage[style=numeric,sorting=none]{biblatex} 
\usepackage{adjustbox}
\usepackage{amsmath}
\usepackage{multirow}
\usepackage{color}
\usepackage{algorithm}
\usepackage{algorithmic}
\usepackage{setspace}
\usepackage{wrapfig}
\usepackage{enumitem}
\usepackage{pifont} % for checkmark and cross
\setlist{topsep=0pt, leftmargin=*}
\usepackage{graphicx}
\usepackage{caption}
\usepackage{subcaption}
\usepackage{hyperref}       % hyperlinks

\newtheorem{theorem}{Theorem}
\newtheorem{lemma}[theorem]{Lemma}
\newtheorem{proposition}[theorem]{Proposition}

\title{Multi-Alignment Contrastive Learning for Enzyme--Reaction Retrieval}

\author{%
Gengmo Zhou$^{1,2}$, \ Feng Yu$^2$,\ Wenda Wang$^1$, \ Zhifeng Gao$^2$, \ Guolin Ke$^2$ , \\
\textbf{Zhewei Wei$^1$}\thanks{\scriptsize Corresponding authors.}, \ \textbf{Zhen Wang$^2$}\footnotemark[1]  \\
$^1$Renmin University of China \quad $^2$DP Technology \\
\texttt{\{zgm2015, wangwenda87, zhewei\}@ruc.edu.cn}, \\
\texttt{\{wangz, yufeng, kegl, gaozf\}@dp.tech}
}

\begin{document}

\maketitle

\input{sections/abstract}
\input{sections/introduction}
\input{sections/related_work}
\input{sections/preliminaries}
\input{sections/method}
\input{sections/experiment}
\input{sections/conclusion}

% \begin{ack}
% Use unnumbered first level headings for the acknowledgments. All acknowledgments

% Do {\bf not} include this section in the anonymized submission, only in the final paper. You can use the \texttt{ack} environment provided in the style file to automatically hide this section in the anonymized submission.
% \end{ack}

{
\small
\printbibliography
}

%%%%%%%%%%%%%%%%%%%%%%%%%%%%%%%%%%%%%%%%%%%%%%%%%%%%%%%%%%%%
\newpage
\appendix
\input{sections/appendix}

%%%%%%%%%%%%%%%%%%%%%%%%%%%%%%%%%%%%%%%%%%%%%%%%%%%%%%%%%%%%

\end{document}

%% file: sections/abstract.tex
\begin{abstract}
Identifying enzymes that catalyze target biochemical reactions is a key step in computational enzyme discovery and biocatalyst design. Recent representation-learning methods formulate this problem as enzyme--reaction matching, where paired enzymes and reactions are embedded into a shared space. However, most existing approaches primarily rely on pairwise enzyme--reaction supervision and make limited use of the relationships within reaction sets or enzyme families. This work introduces a multi-alignment contrastive learning framework for biochemical retrieval. The framework jointly models cross-domain compatibility between enzymes and reactions and within-domain relationships induced by functional annotations. In addition, a Gromov--Wasserstein-inspired regularization objective encourages geometric consistency between the learned enzyme and reaction representation spaces. By combining pairwise catalytic supervision with higher-order relational alignment, the model captures both direct enzyme--reaction associations and broader functional organization. We evaluate the approach on enzyme virtual screening and bidirectional enzyme--reaction retrieval tasks. Experiments on EnzymeMap show improved early-recognition performance under BEDROC and enrichment-factor metrics compared with strong contrastive baselines. On ReactZyme, the method achieves consistent gains across time-based, enzyme-similarity, and reaction-similarity splits, demonstrating robustness to unseen enzymes and unseen reactions. Ablation studies further indicate that within-domain alignment, functional supervision, and the geometric regularization term each contribute to the observed improvements. These results suggest that modeling multiple forms of alignment can improve contrastive retrieval models for enzyme discovery, reaction annotation, and related computational biology applications.
\end{abstract}

%% file: sections/introduction.tex
\section{Introduction}
Enzymes play a vital role in various biological processes, such as biosynthesis and metabolism. They act as catalysts, speeding up chemical reactions in living organisms. However, in the vast protein sequence databases, such as UniProt \cite{uniprot2025uniprot}, only about 1/5 of proteins have been experimentally verified, and only 0.23\% have received sufficient attention from researchers \cite{ribeiro2023enzyme}. Many potentially valuable enzymes may remain hidden among billions of unexplored sequences.

Traditional computational methods, including those based on sequence similarity \cite{altschul1990basic, desai2011modenza, altschul1997gapped}, homology \cite{krogh1994hidden, steinegger2019hh}, and structure \cite{roy2012cofactor, zhang2017cofactor}, often require substantial manual effort and computational resources. These methods also suffer from protein annotation errors.
In recent years, machine learning methods have shown great promise for enzyme screening and retrieval. CLEAN \cite{yu2023enzyme} improved the assignment of EC numbers to enzymes. CLIPZyme \cite{mikhael2024clipzyme} employs contrastive learning to align enzyme and reaction representations for in-silico enzyme screening. ReactZyme \cite{hua2024reactzyme} introduces the largest enzyme-reaction benchmark to date. CARE\cite{yang2024care} integrates textual descriptions of enzymes and reactions to enhance performance. While these methods emphasize enzyme–reaction interactions, they largely overlook the hierarchical complexity inherent in each domain.

In this work, we propose FGW-CLIP, a novel contrastive learning framework based on the optimization of fused Gromov-Wasserstein distance. This framework incorporates multiple alignment strategies, including cross-domain representation alignment between reactions and enzymes, as well as intra-domain alignment within enzyme and reaction representations. By introducing tailored regularization terms, FGW-CLIP minimizes the Gromov-Wasserstein distance between the enzyme and reaction spaces during training, thereby promoting more effective information interaction across both domains.

We provide theoretical insights into FGW-CLIP from the perspective of optimizing the fused Gromov-Wasserstein distance. Furthermore, we offer empirical support by validating our approach on the widely-used benchmark EnzymeMap and ReactZyme for enzyme virtual screening. FGW-CLIP outperforms existing baselines, achieving state-of-the-art performance on both benchmarks.

Our key contributions are as follows:

\begin{itemize}
\item We introduce FGW-CLIP, 
a contrastive learning framework that, to the best of our knowledge, is the first to incorporate fused Gromov-Wasserstein (FGW) distance into enzyme–reaction representation learning. Unlike prior approaches that focus only on inter-domain alignment, FGW-CLIP jointly models both inter-domain and intra-domain alignment, effectively capturing the interplay between reaction and enzyme domains as well as their internal structure.

\item Our formulation integrates FGW into the inverse optimal transport (IOT) framework. This represents a novel theoretical extension of IOT to multi-domain biochemical data, for which we are not aware of any prior work.

\item FGW-CLIP achieves consistent improvements over strong baselines on EnzymeMap and ReactZyme. Ablation studies further highlight the importance of GW-based alignment in preserving intra-domain structural information and improving overall retrieval accuracy.

\end{itemize}

%% file: sections/related_work.tex
\section{Related Work}
\subsection{Contrastive Learning}
Contrastive learning has found extensive applications in vision and multimodal representation learning. CLIP (Contrastive Language-Image Pretraining)~\cite{radford2021learning} advances this paradigm by aligning image and text representations, enabling strong performance in tasks such as image classification, text generation, and human-computer interaction. MLIP \cite{zhang2024mlip} improves CLIP by integrating spatial and frequency-domain information for more comprehensive multimodal understanding. iCLIP \cite{wei2023iclip} further bridges contrastive learning and image classification by optimizing CLIP for both visual tasks and vision-language pairings. X-MoRe \cite{eom2023cross} refines CLIP embeddings to enhance performance in image-to-text and text-to-image retrieval tasks, improving its adaptability for real-world applications.

\subsection{Fused Gromov-Wasserstein Distance}
The Gromov-Wasserstein (GW) Distance is a metric from optimal transport theory that measures the similarity between two metric spaces by considering the structures of the spaces rather than their individual points. Mémoli \cite{memoli2011gromov, scetbon2022linear} proves that $GW^{1/2}$ defines a distance on the space of metric measure spaces quotiented by measure-preserving isometries. Fused Gromov-Wasserstein (FGW) distance \cite{titouan2019optimal,ma2024fused} extends the GW metric to calculate transportation distance between two unregistered probability distributions on different product metric spaces, such as combining graph signals and structures, making it suitable for attributed graphs.

\subsection{Enzyme Screening}
Enzyme virtual screening and recognition accelerate the discovery of new enzymes and drug candidates by accurately identifying functions and efficiently screening potential inhibitors from large libraries. CLIPZyme \cite{mikhael2024clipzyme} is a contrastive learning framework that effectively encodes and aligns representations of enzymes and their corresponding reaction pairs for in-silico enzyme screening. ReactZyme \cite{hua2024reactzyme} introduces a large-scale benchmark for enzyme-reaction prediction, enabling improved enzyme retrieval and function annotation. CARE \cite{yang2024care} incorporates textual descriptions of enzymes and reactions, and provides a benchmark suite for enzyme classification and retrieval.
CLEAN \cite{yu2023enzyme} improves EC number assignment by accurately annotating understudied enzymes and correcting mislabeled entries. In addition, traditional sequence-similarity-based tools and machine learning models, such as BLASTp \cite{altschul1990basic}, DeepEC \cite{wang2020deepec} and ProteInfer \cite{sanderson2023proteinfer}, can also be used for EC number prediction.

%% file: sections/preliminaries.tex
\section{Preliminaries}
To provide a concise understanding of the mathematical foundations underlying our method, we briefly introduce the key concepts of Optimal Transport (OT), Gromov-Wasserstein (GW) distance, and their fused variant FGW.

Optimal Transport (OT) provides a powerful framework for comparing probability distributions by minimizing the cost of transporting mass between them. The Wasserstein distance, a classical OT metric, quantifies the minimal transportation cost between two distributions when their elements reside in a shared metric space. Given two discrete probability distributions $P \in \mathbb{R}^n$ and $Q \in \mathbb{R}^m$, and a non-negative cost matrix $M \in \mathbb{R}^{n \times m}$, the Wasserstein distance is defined as:
\begin{equation}
W(p, q) = \min_{\Gamma \in \tau(p, q)} \langle \Gamma, M \rangle
\end{equation}
where $\langle \cdot, \cdot \rangle$ denotes the Frobenius inner product, and the feasible set $\tau(p, q)$ is given by:
\begin{equation}
\tau(p, q) = \left\{ \Gamma \in \mathbb{R}_+^{n \times m} \mid \Gamma \mathbf{1}_m = P,\ \Gamma^\top \mathbf{1}_n = Q \right\}
\end{equation}
Here, $\mathbf{1}_m$ and $\mathbf{1}_n$ are vectors of all ones. In practice, $P$ and $Q$ are often set to uniform distributions due to their simplicity.

While Wasserstein distance requires both distributions to reside in the same space, Gromov-Wasserstein (GW) distance extends OT to the comparison of structured data that lie in different metric spaces. It aligns relational structures rather than feature correspondences, making it particularly useful for graphs, point clouds, or molecular structures. 
Formally, let \( (X_1, C^1, p) \) and \( (X_2, C^2, q) \) be two metric measure spaces, where \( C^1 \in \mathbb{R}^{n \times n} \) and \( C^2 \in \mathbb{R}^{m \times m} \) denote the pairwise distance matrices within each domain (e.g., reaction and enzyme domains), and \( p \in \Delta^n, q \in \Delta^m \) represent the marginal distributions over \(X_1\) and \(X_2\), respectively. The GW distance is then defined as:
\begin{equation}
\label{eq:gw}
\mathrm{GW}(C^1, C^2, p, q) = \min_{\Gamma \in \tau(p,q)} \sum_{i,j,k,l} \mathcal{L} \left( C^1_{i,k}, C^2_{j,l} \right) \Gamma_{i,j} \Gamma_{k,l}
\end{equation}
where $\mathcal{L} \in \mathbb{R}^{n \times n \times m \times m}$ is typically a squared loss function measuring the discrepancy between intra-domain distances.

Intuitively, the GW distance evaluates how well the intra-domain relational structures are preserved under a soft alignment \(\Gamma\) between entities in the two spaces. Unlike the classical Wasserstein distance, GW focuses on comparing pairwise distances rather than pointwise features, making it suitable for structural matching tasks such as graph alignment.

To combine the geometric alignment of GW with feature-level correspondence, the Fused Gromov-Wasserstein (FGW) distance was introduced. It integrates both ground cost $M$ and relational structures $(C^1, C^2)$ via a fusion parameter $\alpha \in [0,1]$
\begin{equation}
\begin{aligned}
\mathrm{FGW}(p, q, M, C^1, C^2) = \min_{\Gamma \in \tau(p, q)} (1 - \alpha) \langle \Gamma, M \rangle  \\
\quad + \alpha \sum_{i,j,k,l} \mathcal{L}(C^1_{i,k}, C^2_{j,l}) \Gamma_{i,j} \Gamma_{k,l}
\end{aligned}
\end{equation}

The trade-off parameter $\alpha$ balances local feature-level and structural alignment. Empirical studies have demonstrated that FGW effectively captures both content and relational similarities in heterogeneous domains. This makes it a natural choice for applications such as enzyme-reaction screening, where entities (e.g., molecules and proteins) differ both in features and structural topology.

%% file: sections/method.tex
\section{Method}

\subsection{Overview of FGW-CLIP}

\begin{figure*}[t]
    \centering
    \includegraphics[width=\textwidth]{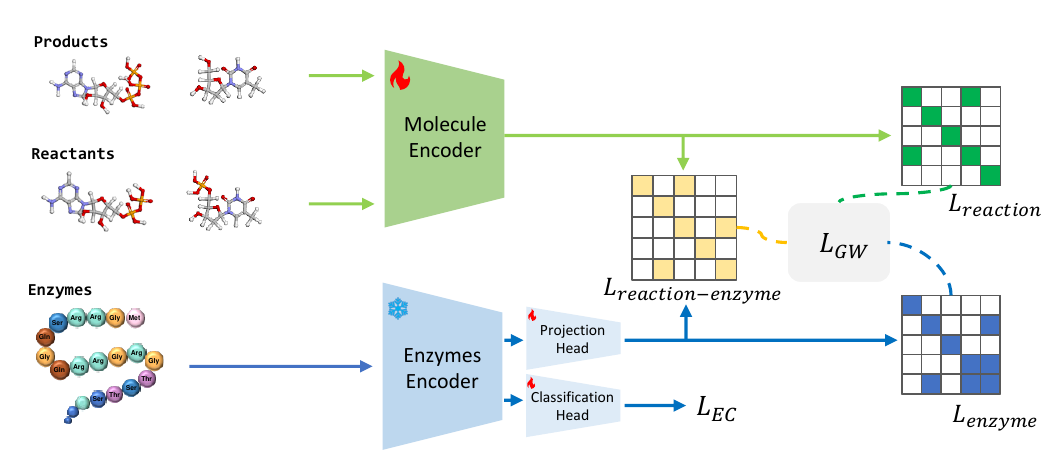}
    \caption{Overview of FGW-CLIP Framework. Reactants and products, along with their structures, are input into the Molecule Encoder, while enzyme sequences are input into the Enzymes Encoder. Representations of enzymes and reactions are aligned using \(L_{\text{reaction-enzyme}}\). Internal alignments within reactions and enzymes are achieved via \(L_{\text{reaction}}\) and \(L_{\text{enzyme}}\), respectively, utilizing EC numbers. Inspired by the goal of minimizing the Gromov-
Wasserstein (GW) distance, we introduce a regularization term \(L_{\text{GW}}\). During training, the Enzyme Encoder is frozen, and a projection head is appended to its output. For clarity, only one classification head is shown in the figure.}
    \label{fig:overview}
\end{figure*}

Given large protein libraries, enzyme screening aims to identify enzymes that can catalyze specific chemical reactions.
We formulate this task as a dense retrieval problem. Trained encoders generate representations for both reactions and enzymes. Reactions are then used as queries to retrieve enzymes, which are ranked based on their similarity to the query reactions. Enzyme–reaction retrieval is a bidirectional task, where reactions can be used to retrieve enzymes and enzymes can also be used to retrieve reactions. The top-ranked enzymes or reactions with the highest similarity scores are considered the most likely candidates for catalysis or functional annotation.

We propose FGW-CLIP from the perspective of optimizing the fused Gromov-Wasserstein distance. As shown in Figure 1, we use the pretrained 3D molecular model Uni-Mol~\cite{zhou2023unimol} to encode the molecules involved in a reaction, and obtain the reaction representation via a readout function. For enzymes, we adopt the pretrained protein language model ESM2~\cite{lin2023evolutionary} to extract sequence embeddings. The ESM2 backbone is kept frozen during FGW-CLIP training, and a linear projection head is added on top to map the embeddings into a shared representation space.
FGW-CLIP performs contrastive learning between reactions and enzymes, where each reaction–enzyme pair is labeled as positive if the enzyme can catalyze the reaction, and negative otherwise. In addition, we incorporate the intrinsic connections within enzymes and reactions by leveraging the Enzyme Commission (EC) number. Each reaction or enzyme is associated with one or more EC numbers, which we store as a list per data point. Taking reactions as an example, if a reaction’s EC number appears in another sample’s EC list within the batch, they are treated as a positive pair; otherwise, as a negative pair. To better align the structural relationships between and within the reaction and enzyme spaces, we introduce a novel regularization term that minimizes the GW distance between them. Furthermore, we integrate an auxiliary EC number prediction loss into the FGW-CLIP framework, enabling effective functional classification.

In the following sections, we elaborate on the core components of FGW-CLIP. Section~\ref{sec:encoder} introduces the molecular encoder Uni-Mol and enzyme encoder ESM2. Section~\ref{sec:Training Strategy} details the training strategy of FGW-CLIP, focusing on contrastive learning between reactions and enzymes. Section~\ref{sec:fgw} provides a theoretical insight into FGW-CLIP from the perspective of FGW distance optimization.

\subsection{Pretraining Backbone of Reaction and Enzyme}
\label{sec:encoder}
For reaction representation, we employ Uni-Mol \cite{zhou2023unimol} to encode the molecules involved in reactions. Uni-Mol is pretrained on large-scale molecular datasets with 3D conformations, which helps integrate 3D structural information. We obtain the reaction embedding using a sum-based readout function.

For enzyme representation, we adopt ESM2 \cite{lin2023evolutionary}, a protein language model pretrained on millions of protein sequences. In our framework, we leverage ESM2 to encode enzyme sequences by averaging all residue embeddings to obtain a single protein-level representation. ESM2 is kept frozen during training.

\subsection{Training Strategy of FGW-CLIP}
\label{sec:Training Strategy} 

\subsubsection{Contrastive Learning Objectives}
We adopt contrastive learning to align reaction and enzyme representations based on experimentally validated catalytic relationships provided in the data. Each training sample consists of a reaction, its corresponding enzyme, and an associated EC number. Since the relationship between reactions and enzymes is not strictly one-to-one, we construct a ground truth label matrix from the data. If a reaction can be catalyzed by a specific enzyme, the corresponding entry in the matrix is set to 1; otherwise, it is 0.

Let \(R_i\) denote the index set of enzymes that can catalyze reaction \(i\), and let \(E_i\) denote the index set of reactions catalyzed by enzyme \(i\). The inter-domain contrastive loss is defined using the InfoNCE \cite{oord2018representation} formulation: 

\begin{equation}  
\label{eq lre}
L_{\text{reaction-enzyme}} = \frac{1}{2} \sum_{i=1}^{N} \left[ 
 -\sum_{j \in R_{i}}\log \frac{e^{(\text{sim}(r_i, e_j)/\tau)}}{\sum_{k=1}^{N} e^{(\text{sim}(r_i, e_k)/\tau)}}
 -\sum_{j \in E_{i}}\log \frac{e^{(\text{sim}(e_i, r_j)/\tau)}}{\sum_{k=1}^{N} e^{(\text{sim}(e_i, r_k)/\tau)}} \right]
\end{equation}
% sim
Here, \(r_i\) and \(e_i\) denote the learned representations of reaction \(i\) and enzyme \(i\), respectively. \(\text{sim}(\cdot, \cdot)\) represents the similarity function, and \(\tau\) is the temperature parameter. We use cosine similarity as \(\text{sim}(\cdot, \cdot)\) throughout our experiments.

In addition to modeling the catalytic relationship between reactions and enzymes, we also incorporate the internal relationships within enzymes and reactions. Let \(I_{i}\) denote the set of samples that share the same EC number with item \(i\). The internal contrastive loss is given as:

\begin{equation}  
\label{eq interbal}  
L_{\text{internal}} = -\sum_{i=1}^{N}\sum_{j \in I_{i}}\log \frac{e^{(\text{sim}(x_i, x_j)/\tau)}}{\sum_{k=1}^{N} e^{(\text{sim}(x_i, x_k)/\tau)}}
\end{equation}

Here, \( x_i \) and \( x_j \) refer to either reactions or enzymes embeddings, while \( x_k \) includes all possible embeddings in the same batch.

\subsubsection{EC Prediction}

The Enzyme Commission (EC) number is a standardized numerical classification scheme for enzymes based on the chemical reactions they catalyze. Each EC number consists of four levels (e.g., EC 1.1.1.1), where each level adds increasing specificity to the enzyme’s function. In our framework, we use enzyme representations to predict EC classes across all four hierarchical levels. For each level, a separate classification head is employed, and we use cross-entropy loss for the predictions. The total EC classification loss is the sum of the losses from all four levels:

\begin{equation}  
\label{eq EC}  
L_{\text{EC}} = L_{\text{level 1}} + L_{\text{level 2}} + L_{\text{level 3}} + L_{\text{level 4}}
\end{equation}

This multi-level classification framework captures the hierarchical nature of enzyme functions and enhances the biological relevance of the learned representations.

\subsubsection{Regularization Loss for GW Distance Optimization}

To enhance the alignment between the reaction and the enzyme spaces, we introduce an additional regularization term motivated by the goal of minimizing the Gromov-Wasserstein (GW) distance. 

This regularization is essential to maintain the internal structure of both spaces while aligning them effectively. The complete loss function is given as:
\begin{equation}
\label{eq GW}
L_{\text{GW}} = - \sum\limits_{i,j,i',j'=1}^{N} \Gamma_d^{\psi_1}(i,i') \Gamma_d^{\psi_2}(j,j') \Gamma^\theta(i,j) \Gamma^\theta(i',j')
\end{equation}

Here, \(\psi_1\) and \(\psi_2\) denote the reaction and enzyme spaces, respectively. The terms \(\Gamma_d^{\psi_1}(i,i')\) and \(\Gamma_d^{\psi_2}(j,j')\) represent the pairwise structural relationships within the reaction and enzyme spaces. These matrices are computed based on intra-domain similarity. They are detached during training to block gradient flow from propagating and thus avoid interference with the optimization of other terms. \(\Gamma^\theta(i,j)\) models the soft alignment between reaction \(i\) and enzyme \(j\), computed from the similarity between their representations. The corresponding formulations are:

\begin{equation}
\label{eq gamma}
\Gamma_d^{\psi_1}(i,i') = \frac{e^{(\text{sim}_{d}(r_i, r_{i'})/\tau)}}{\sum_{k=1}^{N} e^{(\text{sim}_{d}(r_i, r_k)/\tau)}}, \quad
\Gamma^\theta(i,j) = \frac{e^{(\text{sim}(r_i, e_j)/\tau)}}{\sum_{j'=1}^{N} e^{(\text{sim}(r_i, e_{j'})/\tau)}}
\end{equation}

\(\Gamma_d^{\psi_2}(j,j')\) is defined analogously using enzyme representations. This objective leverages the internal structural relationships within both domains to guide inter-domain alignment. A full derivation of this regularization term within the FGW-CLIP framework is provided in Appendix~\ref{app:FGW-CLIP-Framework}, and its computational complexity is discussed in Appendix~\ref{app:GW}.

\subsection{FGW-CLIP: Enhancing CLIP by Optimizing Gromov-Wasserstein Distance}
\label{sec:fgw}
By integrating the training objectives in Section \ref{sec:Training Strategy}, we can derive the overall training objective for FGW-CLIP, denoted as \(L_{\text{FGW}}\), as follows:

\begin{equation}
\label{eq fgw}
L_{\text{FGW}} =  (1-\alpha)(L_{\text{reaction-enzyme}} + L_{\text{reaction}} + L_{\text{enzyme}})
 - 2\alpha L_{\text{GW}} + \lambda L_{\text{EC}}
\end{equation}

The loss function involves multiple terms. Based on the approach outlined in \cite{shi2023understanding, zhou2024smolsearch}, we establish a connection between \(L_{FGW}\) and the fused Gromov-Wasserstein distance optimization problem under a specific constraint through the proposition \ref{FGW-CLIP-Framework}.

\begin{proposition}
\label{FGW-CLIP-Framework}
Given encoder \(f_{\psi_{1}}\) for data field \(X_{1}\) and encoder \(f_{\psi_{2}}\) for data field \(X_{2}\), \(x_{\psi_{1}}\) represents the l2 normalized embeddings of  \(X_{1}\) from \(f_{\psi_{1}}\), while \(x_{\psi_{2}}\) represents the l2 normalized embeddings of \(X_{2}\) from \(f_{\psi_{2}}\). \(\Gamma^{f_{1}}\) represents the label on  \(X_{1}\), \(\Gamma^{f_{2}}\) represents the label on \(X_{2}\), \( \Gamma^{cor}\) represents the label on the pairs \((X_{1}, X_{2})\). FGW-CLIP could be derived from optimizing a specific constraint-fused Gromov-Wasserstein distance as follows:
\begin{equation}  
\label{FGW-CLIP_framework}  
\begin{aligned}  
& \min_{\theta, \psi_{1}, \psi_{2}} \Big\{ (1-\alpha)KL(\Gamma^{cor}||\Gamma^{\theta}) + \alpha GW(\Gamma^{\psi_{1}}_d, \Gamma^{\psi_{2}}_d, \Gamma^{\theta}) \\
& \quad + \lambda_{1} KL(\Gamma^{f_{1}}||\Gamma^{\psi_{1}}) + \lambda_{2} KL(\Gamma^{f_{2}}||\Gamma^{\psi_{2}}) \\ 
& - \lambda_{ce} CE(y_{\psi_{1}}, f_{\psi_{1}}(X_{1})) \Big\} \\
& \text{subject to} \quad   
\begin{aligned}[t]  
    & \Gamma^{\theta} = \arg\min_{\Gamma \in U(a^{cor})} \left( \langle C^{\theta}, \Gamma \rangle - \tau H(\Gamma) \right), \\
    & \Gamma^{\psi_{1}} = \arg\min_{\Gamma \in U(a^{\psi_{1}})} \left( \langle C^{\psi_{1}}, \Gamma \rangle - \tau H(\Gamma) \right), \\
    & \Gamma^{\psi_{2}} = \arg\min_{\Gamma \in U(a^{\psi_{2}})} \left( \langle C^{\psi_{2}}, \Gamma \rangle - \tau H(\Gamma) \right), \\
\end{aligned}  
\end{aligned}  
\end{equation}

where \( KL(X||Y) = \sum\limits_{ij}x_{ij}log\frac{x_{ij}}{y_{ij}} - x_{ij} + y_{ij}\) represents the Kullback-Leibler divergence, and \(H(\Gamma) = -\sum\limits_{i,j} \Gamma_{ij}(\log(\Gamma_{ij}) - 1)\) represents entropic regularization. \( \Gamma^{\theta}, \Gamma^{\psi_{1}}, \Gamma^{\psi_{2}} \in R_{+}^{N \times N}\),
\(C^{\theta}, C^{\psi_{1}}, C^{\psi_{2}} \in R_{+}^{N \times N}\)  are cost matrix and \(C^{\theta}(i,j) = c -x_{\psi_{1},i}x_{\psi_{2},j}^{T}\), \(C^{\psi_{1}}(i,j) = c -x_{\psi_{1},i}x_{\psi_{1},j}^{T}\), \(C^{\psi_{2}}(i,j) = c -x_{\psi_{2},i}x_{\psi_{2},j}^{T}\).  \(a^{\psi_{1}}, a^{\psi_{2}}, a^{cor} \) represent the label vector of dataset \(X_{1}\), \(X_{2}\) and pair dataset \((X_{1}, X_{2})\).  \(\Gamma^{\psi_{1}}_d, \Gamma^{\psi_{2}}_d\) represent the values of  \(\Gamma^{\psi_{1}}, \Gamma^{\psi_{2}}\) respectively, with the gradients detached, \( GW(\Gamma^{\psi_{1}}_d, \Gamma^{\psi_{2}}_d, \Gamma^{\theta}) = \sum\limits_{i,j=1}^{n}\sum\limits_{i',j' =1}^{n}|\Gamma^{\psi_{1}}_d(i,i') - \Gamma^{\psi_{2}}_d(j,j')|^{2}\Gamma^{\theta}(i,j)\Gamma^{\theta}(i',j')\) is the Gromov-Wasserstein distance. \(CE\) is the cross-entropy loss of data field \(X_{1}\), which is added to facilitate a specific classification task as a regularization term.
\end{proposition}

The proof is provided in the Appendix \ref{app:FGW-CLIP-Framework}.  In \(L_{FGW}\), we utilize \(\Gamma_{\psi_{1}}\) and 
\(\Gamma_{\psi_{2}}\) to learn the structural information of two data domains 
\(X_{1}\) and \(X_{2}\), respectively. Through the optimization of the Gromov-Wasserstein distance, structural alignment at the domain level is achieved. We consider this overall structural alignment information as a supplement and enhancement to the existing label alignment information between the two domains \(X_{1}\) and \(X_{2}\). By optimizing this fused Gromov-Wasserstein distance, we can better extend the generalization capability of the CLIP model and alleviate the issue of insufficient effective labels between domains \(X_{1}\) and \(X_{2}\).

As illustrated in Equ.~\ref{eq fgw}, the training of FGW-CLIP involves the joint optimization of multiple objective functions. The choice of weighting coefficients ($\alpha$ and $\lambda$) affects model performance. We find that smaller values of $\alpha$ reduce the impact of the GW loss, leading to weaker alignment, while larger $\alpha$ values overemphasize the GW optimization, interfering with intra-domain alignment. We also analyze the impact of different $\lambda$ values on EC prediction loss. Since the EC loss tends to have a relatively large magnitude, assigning it a high weight can dominate the overall objective and suppress the contributions of other loss terms, leading to performance degradation. Therefore, we use a smaller weight of 0.1 to maintain a balanced scale between the EC loss and the other components.

%% file: sections/experiment.tex
\section{Experiment}
\subsection{Enzyme Screening}

\subsubsection{Datasets}

\paragraph{EnzymeMap} Based on the original EnzymeMap dataset \cite{D3SC02048G}, it involves biochemical reactions linked to UniProt IDs and EC numbers. 
There are 46,356 enzyme-driven reactions with 16,776 unique chemical reactions, 12,749 enzymes, 2,841 EC numbers, and 394 reaction rules in the EnzymeMap dataset. We split the dataset into training, validation, and test sets based on the reaction rule IDs, with a ratio of 0.8/0.1/0.1, containing 34,427, 7,287, and 4,642 entries, respectively, the same as in CLIPZyme.

\paragraph{Enzyme Screening Set} This dataset integrated the EnzymeMap dataset, Brenda release 2022\_2 \cite{2020BRENDA}, and UniProt release 2022\_01 \cite{2023Enzyme}, and filtered out the sequences that are longer than 650 amino acids. It includes a total of 261,907 protein sequences. Enzyme screening Set is used as a virtual screening database, where we use the reactions from the EnzymeMap test set as queries to perform screening in it.

\subsubsection{Baseline}

In this task, we adopt the previous state-of-the-art method CLIPZyme\cite{mikhael2024clipzyme} and its variants as the baselines. CLIPZyme is a contrastive learning approach for enzyme screening. We follow the experimental setup of CLIPZyme and use the same datasets.

\subsubsection{Evaluation Metric} For this task, we utilize the BEDROC (Boltzmann-Enhanced Discrimination of ROC) \cite{truchon2007evaluating} score and the enrichment factor (EF) as evaluation metrics, which measure early retrieval performance and the concentration of true positives among top-ranked results, respectively. Higher values indicate better performance for both metrics. We calculate BEDROC at $\alpha=85$ and $\alpha=20$, and focus on EF in the top 5\% and 10\% of the predictions, to align with the evaluation protocol in CLIPZyme. Their definitions are in appendix~\ref{app:metrics}.

\subsubsection{Results}

\begin{table}[!ht]
    \centering
    \caption{Enzyme virtual screening performance on EnzymeMap. The higher the BEDROC and EF, the better.}
     \label{tab:main}
    \begin{tabular}{c|cccc}
    \toprule
        Method & $\text{BEDROC}_{85}$(\%) & $\text{BEDROC}_{20}$(\%) & $\text{EF}_{0.05}$ &  $ \text{EF}_{0.1}$ \\ 
    \midrule
        
        CLIPZyme (ESM) & 36.91 & 53.04 & 11.93 & 6.84 \\ 
        CLIPZyme (CGR) & 38.91 & 57.58 & 13.16 & 7.73 \\
        CLIPZyme & 44.69 & 62.98 & 14.09 & 8.06 \\ 
        FGW-CLIP & \textbf{48.66} & \textbf{66.69} & \textbf{14.91} & \textbf{8.18} \\ 
    \bottomrule
    \end{tabular}
\end{table}

Table \ref{tab:main} shows the performance comparison between FGW-CLIP and the current SOTA baseline CLIPZyme on EnzymeMap. The best results for each metric are shown in bold. We also include variants of CLIPZyme using different combinations of protein and reaction encoders. ESM denotes using the ESM2 protein language model for enzyme representation. CGR \cite{hoonakker2011condensed} is a method for obtaining reaction representations based on graph structures. These results are consistent with those reported in the original CLIPZyme paper. As shown in the table, FGW-CLIP achieves the best performance across all four metrics for BEDROC and EF. Notably, it achieves a substantial improvement of approximately 4\% on both BEDROC scores. This demonstrates the advantage of FGW-CLIP in optimizing the fused GW distance through contrastive learning, which jointly captures inter-domain alignment between enzymes and reactions as well as intra-domain structural relationships.

\begin{table}[ht]
\centering
\caption{Performance on EnzymeMap. Exclude enzymes that appeared in the training set from the screening set.}
 \label{tab:gen}
\begin{tabular}{c|cccc}
\toprule
 Method & $\text{BEDROC}_{85}$(\%) & $\text{BEDROC}_{20}$(\%) & $\text{EF}_{0.05}$ &  $ \text{EF}_{0.1}$ \\ 
\midrule
 Clipzyme  &  39.13	& 58.86  &  13.40 & \textbf{7.81} \\ 
 FGW-CLIP &  \textbf{45.14}	& \textbf{61.43}  &  \textbf{13.57} & 7.61 \\ 

\bottomrule
\end{tabular}
\end{table}

We further evaluate the generalization ability of FGW-CLIP by testing on unseen enzymes. Specifically, we exclude all enzymes in the screening set that also appeared in the training set. Table \ref{tab:gen} presents the results, showing that FGW-CLIP outperforms CLIPZyme on 3 out of the 4 evaluation metrics, and performs comparably on EF$_{0.1}$. Notably, FGW-CLIP achieves a substantial improvement in both BEDROC scores, demonstrating its ability to identify catalytically relevant enzymes even without prior exposure. These results highlight FGW-CLIP’s stronger capacity to capture generalizable features of enzyme–reaction interactions.

\subsection{Enzyme-Reaction Retrieval}

\begin{table*}[ht]
\centering

\caption{Retrieval Performance on ReactZyme of the time-based split. 
The best results are \textbf{bolded}. Except for the mean rank, the higher the metric, the better. The baselines with * indicate that they use FANN \cite{puny2021frame} to enhance residue-level representations.
}
\label{tab:time.split}
\subfloat[Given the enzyme, retrieve positive reactions.]{
\resizebox{\columnwidth}{!}{%
\small
\begin{tabular}{l|cccc|cccc|c|c}
\toprule
Time/enzyme-reaction & Top1 & Top5 & Top10 & Top20 & Top1-N & Top5-N & Top10-N & Top20-N & Mean Rank & MRR \\
\midrule
Ground-truth & 1.0000 & 1.0000 & 1.0000 & 1.0000 & 1.0000 & 0.2002 & 0.1001 & 0.0500 & 1.0004 & 0.9998 \\
\midrule
MAT-2D + ESM & 0.3246 & 0.6044 & 0.7079 & 0.7972 & 0.3246 & 0.1209 & 0.0708 & 0.0399 & 40.4756 & 0.4549 \\
MAT-2D + SaProt & 0.2073 & 0.4020 & 0.5004 & 0.6120 & 0.2073 & 0.0804 & 0.0499 & 0.0306 & 75.3546 & 0.2898 \\
UniMol-2D + ESM & 0.2827 & 0.5210 & 0.6508 & 0.7612 & 0.2827 & 0.1041 & 0.0651 & 0.0380 & 53.4261 & 0.4011 \\
UniMol-2D + SaProt & 0.1957 & 0.3855 & 0.4380 & 0.6021 & 0.1957 & 0.0771 & 0.0438 & 0.0301 & 79.8460 & 0.2788 \\
UniMol-2D + ESM* & 0.2948 & 0.5866 & 0.6912 & 0.7831 & 0.2948 & 0.1173 & 0.0691 & 0.0391 & 45.0611 & 0.4289 \\
UniMol-2D + SaProt* & 0.2512 & 0.4329 & 0.6474 & 0.6879 & 0.2512 & 0.0866 & 0.0647 & 0.0344 & 63.1455 & 0.3176 \\
\midrule
MAT-3D + ESM & 0.2858 & 0.4955 & 0.6548 & 0.7405 & 0.2858 & 0.0991 & 0.6550 & 0.0371 & 60.3628 & 0.4041 \\
MAT-3D + SaProt & 0.1210 & 0.2265 & 0.3108 & 0.4015 & 0.1210 & 0.0453 & 0.0311 & 0.0201 & 150.0301 & 0.1862 \\
UniMol-3D + ESM & 0.2905 & 0.5365 & 0.6586 & 0.7639 & 0.2905 & 0.1074 & 0.0659 & 0.0382 & 46.0553 & 0.4104 \\
UniMol-3D + SaProt & 0.0916 & 0.2134 & 0.2923 & 0.3882 & 0.0916 & 0.0426 & 0.0292 & 0.0194 & 168.8244 & 0.1591 \\
UniMol-3D + ESM* & 0.3588 & 0.6545 & 0.7815 & 0.8126 & 0.3588 & 0.1309 & 0.0781 & 0.0406 & 32.7443 & 0.4952 \\
UniMol-3D + SaProt* & 0.2508 & 0.4075 & 0.5448 & 0.6421 & 0.2508 & 0.0815 & 0.0546 & 0.0321 & 59.8345 & 0.3453 \\

\midrule
CLIPZyme (MAT-2D + ESM) & 0.3041 & 0.5993 & 0.6943 & 0.7840 & 0.3041  & 0.1201 & 0.0695 & 0.0392 & 42.3645 & 0.4355 \\
CLIPZyme (UniMol-3D + ESM) & 0.2631 & 0.4534 & 0.6444 & 0.7516 & 0.2631 & 0.0907 & 0.0645 & 0.0376 & 45.3637 & 0.3940 \\
\midrule
FGW-CLIP, EC Mode & \textbf{0.3830} & 0.7090 & 0.8038 & 0.8823 & \textbf{0.3830} & 0.1418 & 0.0804 & \textbf{0.0442} & \textbf{23.1991} & 0.5222 \\

FGW-CLIP, EC Max & 0.3782 & \textbf{0.7143} & \textbf{0.8119} & \textbf{0.8839} & 0.3782 & \textbf{0.1429} & \textbf{0.0812} & 0.0442 & 24.1628 & \textbf{0.5229} \\

    \bottomrule
    \end{tabular}%
}}

\subfloat[Given the reaction, retrieve positive enzymes.]{
\resizebox{\columnwidth}{!}{%
\begin{tabular}{l|cccc|cccc|c|c}
\toprule
Time/reaction-enzyme & Top1 & Top5 & Top10 & Top20 & Top1-N & Top5-N & Top10-N & Top20-N & Mean Rank & MRR \\
\midrule
Data(Ground-truth) & 1.0000 & 1.0000 & 1.0000 & 1.0000 & 1.0000 & 0.4718 & 0.2895 & 0.1677 & 2.8324 & 0.7497 \\
\midrule
MAT-2D + ESM & 0.2175 & 0.3815 & 0.4924 & 0.6033 & 0.2175 & 0.1570 & 0.1206 & 0.0871 & 165.3066 & 0.1789 \\
MAT-2D + SaProt & 0.1260 & 0.2153 & 0.2921 & 0.3778 & 0.1260 & 0.0886 & 0.0716 & 0.0546 & 281.2419 & 0.0981 \\
UniMol-2D + ESM & 0.1435 & 0.2299 & 0.3554 & 0.4367 & 0.1435 & 0.0946 & 0.0871 & 0.0631 & 270.9385 & 0.1233 \\
UniMol-2D + SaProt & 0.0912 & 0.1494 & 0.2252 & 0.3488 & 0.0912 & 0.0615 & 0.0552 & 0.0504 & 536.5624 & 0.0805 \\
UniMol-2D + ESM* & 0.1486 & 0.2294 & 0.3529 & 0.4865 & 0.1486 & 0.0944 & 0.0865 & 0.0703 & 254.1982 & 0.1257 \\
UniMol-2D + SaProt* & 0.0988 & 0.1587 & 0.2273 & 0.3536 & 0.0988 & 0.0653 & 0.0557 & 0.0511 & 504.2854 & 0.0934 \\
\midrule
MAT-3D + ESM & 0.2281 & 0.4240 & 0.5502 & 0.5879 & 0.2281 & 0.1703 & 0.1393 & 0.0852 & 152.1328 & 0.1931 \\
MAT-3D + SaProt & 0.1037 & 0.1800 & 0.2603 & 0.3671 & 0.1037 & 0.0723 & 0.0659 & 0.0532 & 411.5762 & 0.1056 \\
UniMol-3D + ESM & 0.1678 & 0.3155 & 0.3960 & 0.5011 & 0.1678 & 0.1267 & 0.1002 & 0.0748 & 177.4881 & 0.1400 \\
UniMol-3D + SaProt & 0.0558 & 0.0979 & 0.1359 & 0.1918 & 0.0558 & 0.0393 & 0.0344 & 0.0278 & 700.9714 & 0.0538 \\
UniMol-3D + ESM* & 0.2045 & 0.3792 & 0.4475 & 0.5168 & 0.2045 & 0.1523 & 0.1133 & 0.0749 & 167.5862 & 0.1628 \\
UniMol-3D + SaProt* & 0.1331 & 0.2044 & 0.3365 & 0.4119 & 0.1331 & 0.0821 & 0.0852 & 0.0597 & 322.5755 & 0.1122 \\
\midrule
CLIPZyme (MAT-2D + ESM) & 0.1757 & 0.3447 & 0.4555 & 0.5343 & 0.1757 & 0.1312 & 0.1101 & 0.0756 & 173.3521 & 0.1678 \\
CLIPZyme (UniMol-3D + ESM) & 0.1331 & 0.2993 & 0.3554 & 0.4567 & 0.1331 & 0.1033 & 0.0949 & 0.0740 & 186.4576 & 0.1313 \\
\midrule
FGW-CLIP, EC Mode & 0.3918 & 0.6230 & 0.7141 & 0.8060 & 0.3918 & 0.2582 & 0.1845 & 0.1233 & 106.2501 & \textbf{0.3394} \\
FGW-CLIP, EC Max & \textbf{0.3967} & \textbf{0.6317} & \textbf{0.7172} & \textbf{0.8075} & \textbf{0.3967} & \textbf{0.2614} & \textbf{0.1866} & \textbf{0.1235} & \textbf{104.6430} & 0.3392 \\

\bottomrule
    \end{tabular}%
}}
\end{table*}

\subsubsection{Datasets}

\paragraph{ReactZyme} ReactZyme~\cite{hua2024reactzyme} is compiled from the SwissProt and Rhea databases \cite{boeckmann2003swiss, bansal2022rhea}, provides a comprehensive resource for enzyme-reaction prediction. The dataset comprises 178,463 enzyme-reaction pairs, including 178,327 unique enzymes and 7,726 unique reactions. Compared to existing datasets such as ESP \cite{kroll2023general} and EnzymeMap, ReactZyme contains significantly more enzyme-reaction pairs and captures substrate, product, and reaction-level information, though it lacks atom-mapping data.

\paragraph{Data Split} 
Three splits are evaluated, and 10\% of the training data is randomly sampled as the validation set.

Time Split: Based on annotation date, with pairs before 2010-12-31 for training (166,175 pairs) and after for testing (12,287 pairs), ensuring a 93\%/7\% split.

Enzyme Similarity Split: Ensures at least 60\% sequence difference between training and test enzymes, resulting in 169,724 training pairs and 8,739 test pairs (95\%/5\% split).

Reaction Similarity Split: Ensures no overlap in reactions between training and test sets, with 163,771 training pairs and 14,692 test pairs (91\%/9\% split).

\subsubsection{Baselines}

For this task, baseline models include Molecule Attention Transformer-2D (MAT-2D) \cite{maziarka2020molecule} and UniMol-2D \cite{zhou2023unimol} for 2D molecular graphs, and MAT-3D and UniMol-3D for 3D molecular conformations as reaction representations. For enzyme representations, ESM  \cite{lin2023evolutionary} and the structure-aware protein language model SaProt \cite{su2023saprot} are utilized. Additionally, an equivariant graph neural network FANN \cite{puny2021frame} is employed to enhance residue-level representations. For CLIPZyme \cite{mikhael2024clipzyme}, we adopt the implementation provided in the ReactZyme paper~\cite{hua2024reactzyme}. This follows the evaluation protocol defined by ReactZyme to ensure consistency.

\subsubsection{Metrics}

We evaluate performance using Top-k Accuracy, Top-k Accuracy-N, Mean Rank, and Mean Reciprocal Rank (MRR). Top-k Accuracy measures the proportion of cases where the correct enzyme or reaction appears within the top-k predictions, while Top-k Accuracy-N quantifies the average proportion of correct predictions among the top-k results, normalized by k, across all test instances. Mean Rank calculates the average rank of the correct enzyme or reaction, with lower values indicating better performance. MRR assesses the speed of retrieval by averaging the reciprocal ranks of the first correct prediction, where higher values indicate better performance. Their definitions are in appendix~\ref{app:metrics}.

\subsubsection{Results}

We evaluate enzyme and reaction retrieval across the three standard splits of ReactZyme. Results for the time-based split are shown in Table~\ref{tab:time.split}, while results for the enzyme similarity-based and reaction similarity-based splits are provided in Appendix~\ref{app:react}. respectively. The best results in each table are highlighted in bold. Methods marked with * indicate the use of FANN \cite{puny2021frame} for residue-level representation enhancement. Baseline results are reported from the ReactZyme paper.

Due to incomplete EC number annotations in ReactZyme (e.g., containing "-" or tokens like "n11"), we adopt two strategies to impute missing values: Mode filling and Max value filling. 
In Table~\ref{tab:time.split}, these variants are labeled as FGW-CLIP, EC Mode and FGW-CLIP, EC Max. In the Mode variant, missing components are replaced with the most frequent value at the corresponding position in the dataset.
In the Max variant, positions with “n” are filled with the current maximum plus 1, and those with “-” are filled with the maximum plus 2, ensuring uniqueness while preserving dataset integrity. Both strategies enable to construct a complete and consistent dataset for further experiments.

In the experiments based on the time-based split, as shown in Table \ref{tab:time.split}, FGW-CLIP achieves state-of-the-art performance across all metrics. Notably, in the reaction-to-enzyme retrieval task, it outperforms the strongest baseline, with improvements of over 20 percentage in Top-k metrics. 
In the enzyme-similarity-based split, as shown in Table \ref{tab:enzyme.split}], FGW-CLIP consistently surpasses all baselines in both retrieval directions. In the reaction-similarity-based split, as shown in Table \ref{tab:reaction.split}, FGW-CLIP achieves the best performance on all metrics except Mean Rank. 

We note that ReactZyme is highly imbalanced, with 178,327 unique enzymes but only 7,726 unique reactions. Under the reaction-similarity-split, test reactions are entirely disjoint from training data, occasionally resulting in low-ranking hits that affect Mean Rank, while Top-k accuracy remains largely unaffected.

Finally, comparing FGW-CLIP with baselines using the same encoder backbone (UniMol-3D + ESM and CLIPZyme with UniMol-3D + ESM), our method consistently outperforms them across all splits, while maintaining a comparable Mean Rank under the most challenging setting. These results empirically validate the effectiveness and generalization ability of FGW-CLIP.

\subsection{Ablation Study}

\begin{table}[ht]
\centering
\caption{Ablation study of different training strategies in FGW-CLIP on EnzymeMap. “R” represents the reaction, “E” represents the enzyme, and “\(\_\)” indicates the use of contrastive learning between both sides. “EC” represents the addition of an EC prediction head. }
 \label{tab:abla_loss}
\begin{tabular}{c|cccc}
\toprule
   Method & $\text{BEDROC}_{85}$(\%) & $\text{BEDROC}_{20}$(\%) & $\text{EF}_{0.05}$ &  $ \text{EF}_{0.1}$ \\
\midrule
R\_E &  45.94	& 61.11  &  13.28 & 7.41 \\ 
R\_E + R\_R &  48.08	&  63.92 & 13.89 & 7.89 \\ 
% R-E, E-E &  43.18	& 60.34  &  13.44  & 7.65 \\ 
R\_E + EC  &  45.25	&  63.93 & 14.46 & 7.97 \\ 
% R-E, R-R, EC   & 46.70	& 65.25   &  14.72	 &  8.15 \\ 
R\_E + R\_R + E\_E + EC    &  45.83		& 64.17	  &  14.37	& 8.12 \\ 
FGW-CLIP (Line 4 + GW) & \textbf{48.66} & \textbf{66.69} & \textbf{14.91} & \textbf{8.18} \\ 

\bottomrule
\end{tabular}
\end{table}

To understand how various design choices affect the final performance, we conducted comprehensive ablation studies to assess the contributions of different components in FGW-CLIP. First, we evaluate the impact of different training strategies. Table \ref{tab:abla_loss} summarizes the results. 
Comparing R\_E and R\_E + R\_R, we observe that including R\_R, which captures internal relationships within reactions, leads to notable improvements in both BEDROC and EF metrics. This highlights the importance of capturing intra-domain structure for more effective enzyme screening.
Similarly, adding the EC prediction head (R\_E + EC) enhances EF scores compared to the baseline R\_E, suggesting that hierarchical EC information encourages the model to learn biologically meaningful enzyme features.
The full FGW-CLIP framework yields the highest performance across all metrics, confirming the effectiveness of optimizing the Gromov-Wasserstein distance. This demonstrates the benefit of simultaneously modeling inter-domain alignment and preserving internal structural consistency.

\begin{table}[ht]
\centering
\caption{Ablation study of different GW optimization strategies and weight \(\alpha\) on EnzymeMap.}
 \label{abla_fgw}
\begin{tabular}{c|cccc}
\toprule
   Method & $\text{BEDROC}_{85}$(\%) & $\text{BEDROC}_{20}$(\%) & $\text{EF}_{0.05}$ &  $\text{EF}_{0.1}$ \\
\midrule
$\alpha=0.05$ &  45.59	& 63.84  &  14.21 & \textbf{8.24} \\ 
$\alpha=0.3$ &  47.46	&  64.86 & 14.29 & 8.02 \\ 
$\alpha=0.5$ &  47.44	& 64.31  &  14.18  & 7.89 \\ 
Label, $\alpha=0.1$  &  47.08 &  65.28 & 14.64 & 8.08 \\ 
No Detach, $\alpha=0.1$   & 47.63	& 64.96  &  14.39	 &  8.06 \\ 
\midrule
FGW-CLIP (Detach, $\alpha=0.1$) & \textbf{48.66} & \textbf{66.69} & \textbf{14.91} & 8.18 \\ 

\bottomrule
\end{tabular}
\end{table}

\begin{figure*}[ht]
    \centering
    \includegraphics[width=0.43\textwidth]{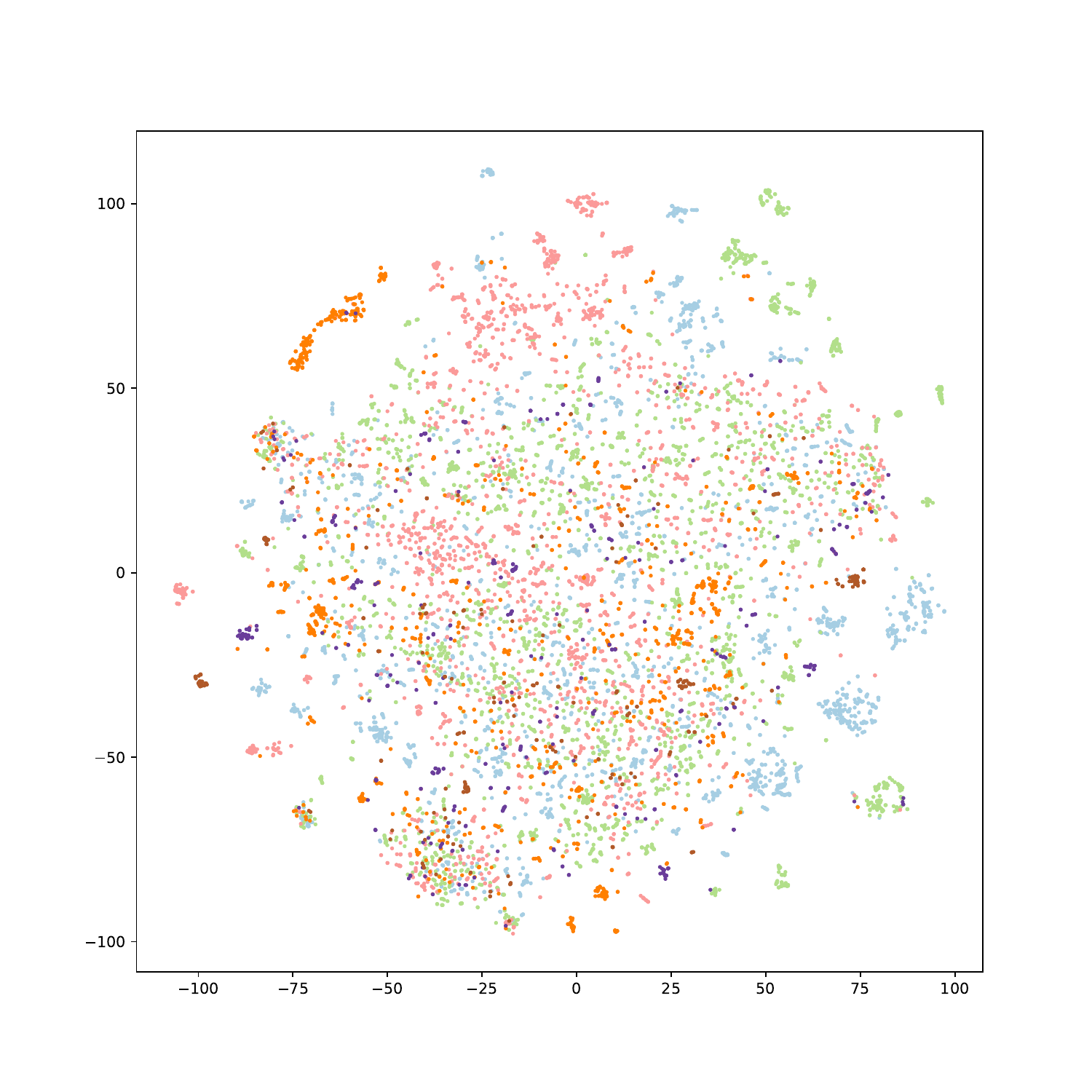}
    \hspace{0.02\textwidth}
    \includegraphics[width=0.43\textwidth]{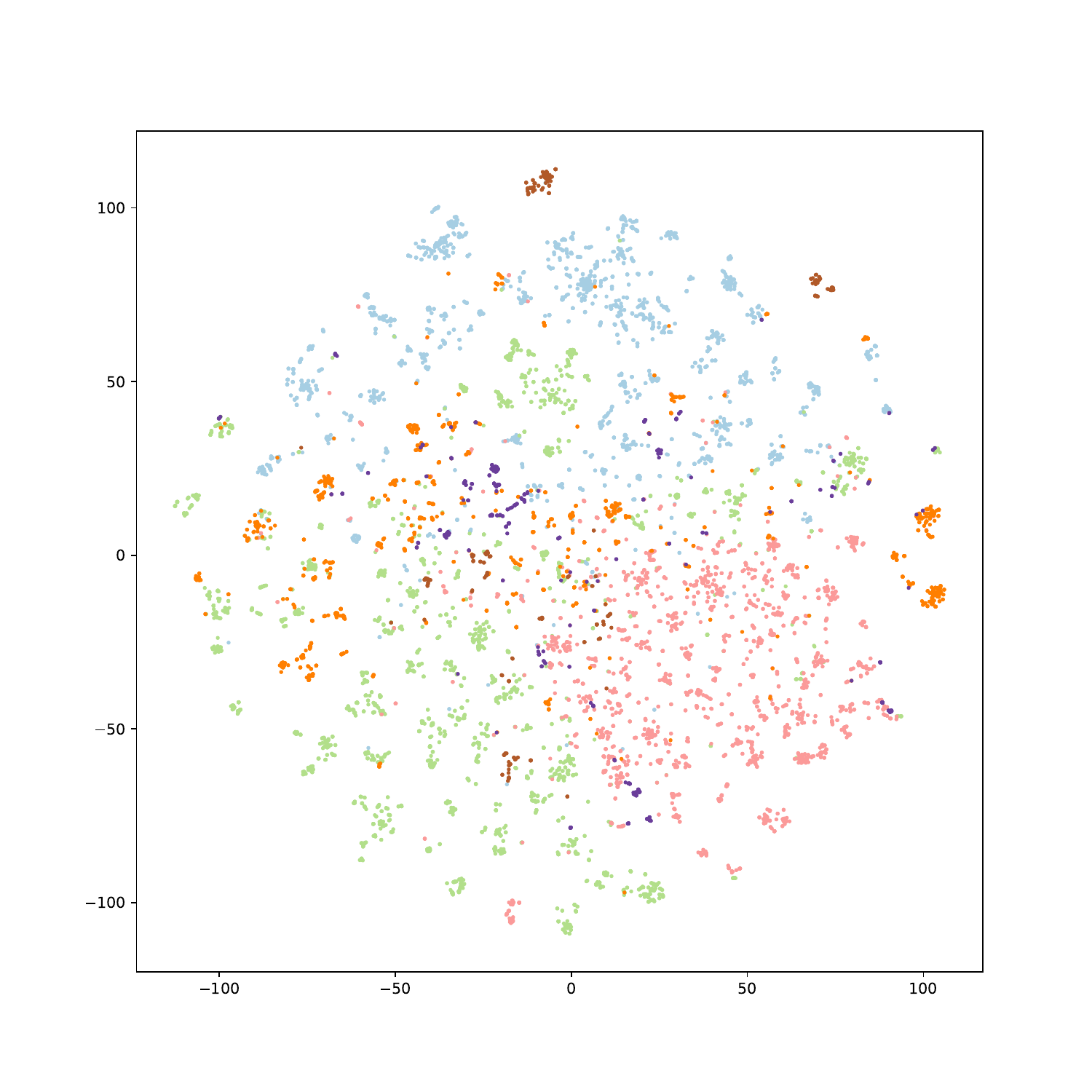}
    \caption{t-SNE visualization of enzyme representations. Left: pretrained checkpoint. Right: FGW-CLIP. Colors represent top-level EC numbers (EC 1 to EC 6).}
    \label{fig:tsne}
\end{figure*}

We further investigate the impact of different GW loss weights \(\alpha\) and incorporation strategies within the FGW-CLIP framework, as summarized in Table~\ref{abla_fgw}.

The parameter \(\alpha\) controls the trade-off between inter-domain alignment and intra-domain structure preservation. We find that \(\alpha = 0.1\) yields the best overall performance.
A smaller value (\(\alpha = 0.05\)) underweights the GW loss, leading to weaker alignment.
In contrast, a larger value (\(\alpha = 0.5\)) overemphasizes GW optimization, which disrupts the intra-domain alignment and impairs the model’s ability to capture meaningful intra-domain structures.

We also compare several GW integration strategies. The “Label” variant replaces the similarity matrices \(\text{sim}_{d}(r_j, r_{j'})\) and \(\text{sim}_{d}(e_i, e_{i'})\) in Eq.~\ref{eq gamma} with binary labels derived from internal contrastive learning (Eq.~\ref{eq interbal}). The “No Detach” variant disables gradient blocking for these matrices, allowing gradients to propagate through them during training.

As shown in Table \ref{abla_fgw}, the default “Detach” strategy, where gradients are not propagated through intra-domain similarity matrices, achieves the best performance. This detachment improves optimization stability by preventing interference between inter-domain alignment and intra-domain structure preservation. This strategy helps strike a meaningful balance, further highlighting the effectiveness of our design.

\begin{table}[ht]
\centering
\caption{Ablation study of EC loss weight \(\lambda\) on EnzymeMap.}
 \label{abla_lambda}
\begin{tabular}{c|cccc}
\toprule
  \(\lambda\) & $\text{BEDROC}_{85}$(\%) & $\text{BEDROC}_{20}$(\%) & $\text{EF}_{0.05}$ &  $\text{EF}_{0.1}$ \\
\midrule
1.0 & 41.93 & 60.10 & 13.46 & 7.68 \\
0.5 & 43.80 & 61.17 & 13.58 & 7.72 \\
0.3 & 45.38 & 64.02 & 14.48 & 8.00 \\ 
\midrule
FGW-CLIP ($\lambda=0.1$) & \textbf{48.66} & \textbf{66.69} & \textbf{14.91} & \textbf{8.18} \\ 

\bottomrule
\end{tabular}
\end{table}

We additionally investigate the impact of the EC prediction loss weight $\lambda$, as shown in Table~\ref{abla_lambda}. Due to its relatively large raw loss magnitude, assigning a high $\lambda$ tends to overshadow the other objectives, which degrades the final performance. As observed, increasing $\lambda$ from 0.3 to 1.0 consistently reduces BEDROC and EF metrics. We therefore choose a small value of 0.1, which ensures that EC supervision is incorporated without dominating the optimization. This highlights the importance of balancing loss magnitudes in multi-objective training.

\subsection{Visualization}

To visually demonstrate the distinctions between embeddings learned by FGW-CLIP and those from pretrained ESM2 checkpoint, we present a comparative visualization in Figure \ref{fig:tsne}. The enzymes depicted are sourced from EnzymeMap, with distinct colors representing different top-level EC numbers ranging from EC 1 to EC 6. Upon comparison, the classification boundaries in Figure \ref{fig:tsne} generated by FGW-CLIP exhibit greater clarity, and the intra-class molecular distances appear more appropriately scaled. Notably, some clusters subdivide into multiple subclusters, potentially reflecting the inherent hierarchical structure within the molecular compositions.

%% file: sections/conclusion.tex
\section{Conclusion}

In this work, we proposed FGW-CLIP, a novel contrastive learning framework that enhances enzyme screening by optimizing the fused Gromov-Wasserstein (FGW) distance. Our approach jointly aligns the reaction and enzyme spaces while preserving the structural relationships within each domain.
FGW-CLIP integrates three contrastive objectives—reaction–enzyme, enzyme–enzyme, and reaction–reaction alignment—along with a Gromov-Wasserstein regularization and an auxiliary EC prediction loss to better capture biochemical relationships.
We conducted extensive experiments on EnzymeMap and ReactZyme, where FGW-CLIP consistently achieved state-of-the-art performance across key metrics such as BEDROC and MRR. Notably, with the same encoder backbone, FGW-CLIP outperformed other baselines on all ReactZyme splits, confirming its robustness and scalability.

In future work, we plan to improve the reaction representation by incorporating reaction directionality, addressing current limitations. Additionally, we will explore the integration of 3D structural information and functional-level optimization, and extend FGW-CLIP to broader biochemical applications such as drug discovery and metabolic pathway prediction.

%% file: sections/appendix.tex
\section{Dataset Details and Baselines}
\label{app:data}
\subsection{Enzyme Screening}

\paragraph{EnzymeMap} Based on the original EnzymeMap dataset \cite{D3SC02048G}, this dataset consists of biochemical reactions linked to UniProt IDs and EC numbers. It contains 46,356 enzyme-driven reactions, including 16,776 unique chemical reactions, 12,749 unique enzymes, 2,841 EC numbers, and 394 reaction rules. Following the setup in CLIPZyme \cite{mikhael2024clipzyme}, we split the dataset into training, validation, and test sets based on reaction rule IDs with a ratio of 0.8/0.1/0.1, resulting in 34,427, 7,287, and 4,642 entries, respectively.

\paragraph{Enzyme Screening Set}  This dataset integrated the EnzymeMap dataset, Brenda release 2022\_2 \cite{2020BRENDA}, and UniProt release 2022\_01 \cite{2023Enzyme}, and filtered out the sequences that are longer than 650 amino acids. The final set includes 261,907 protein sequences. During evaluation, reactions from the EnzymeMap test set are used as queries to screen potential matching enzymes from this pool.

\paragraph{Baselines} In this task, we use the SOTA method CLIPZyme \cite{mikhael2024clipzyme} and several of its variants as baselines. CLIPZyme is a contrastive learning approach for enzyme screening, and we follow its experimental setup.

\subsection{Enzyme-Reaction Retrieval}

\paragraph{ReactZyme} ReactZyme compiled from the SwissProt and Rhea databases \cite{boeckmann2003swiss, bansal2022rhea}, provides a comprehensive resource for enzyme-reaction prediction. The dataset comprises 178,463 enzyme-reaction pairs, including 178,327 unique enzymes and 7,726 unique reactions. Compared to existing datasets such as ESP \cite{kroll2023general} and EnzymeMap, ReactZyme contains significantly more enzyme-reaction pairs and captures substrate, product, and reaction-level information, though it lacks atom-mapping data.

\paragraph{Data Split} ReactZyme provides explicit train-test splits for each setting. Additionally, 10\% of the training set is randomly sampled as a validation set.

\begin{itemize}
    \item Time Split: Based on annotation date, with pairs before 2010-12-31 for training (166,175 pairs) and after for testing (12,287 pairs), ensuring a 93\%/7\% split.
    \item Enzyme Similarity Split: Ensures at least 60\% sequence difference between training and test enzymes, resulting in 169,724 training pairs and 8,739 test pairs (95\%/5\% split).
    \item Reaction Similarity Split: Ensures no overlap in reactions between training and test sets, with 163,771 training pairs and 14,692 test pairs (91\%/9\% split).
\end{itemize}

\paragraph{Baselines} For this task, baseline models include Molecule Attention Transformer-2D (MAT-2D) \cite{maziarka2020molecule} and UniMol-2D \cite{zhou2023unimol} for 2D molecular graphs, and MAT-3D and UniMol-3D for 3D molecular conformations as reaction representations. For enzyme representations, ESM  \cite{lin2023evolutionary} and the structure-aware protein language model SaProt \cite{su2023saprot} are utilized. Additionally, an equivariant graph neural network FANN  \cite{puny2021frame}) is employed to enhance residue-level representations. For CLIPZyme \cite{mikhael2024clipzyme}, we adopt the implementation provided in the ReactZyme paper\cite{hua2024reactzyme}. This follows the evaluation protocol defined by ReactZyme to ensure consistency.

\section{Evaluation metrics}
\label{app:metrics}
\subsection{Enzyme Screening}

Here we introduce the two metrics we use to evaluate the efficiency of the enzyme screening task. The Boltzmann-Enhanced Discrimination of ROC (BEDROC) score is a modified version of the AUC of the ROC curve, which places a stronger emphasis on early enrichment (i.e., at high-ranking positions). This is particularly important in drug discovery, where experimental testing is costly, and being able to identify potentially active compounds early on can save significant time and resources. The calculation formula of BEDROC is shown as follows :

\begin{align}
\text{BEDROC} = & \frac{\sum_{i=1}^n e^{-\alpha r_i / N}}{\frac{R_{\alpha}\left(1-e^{-\alpha}\right)}{e^{\alpha / N}-1}} \times \frac{R_{\alpha} \sinh (\alpha / 2)}{\cosh (\alpha / 2) - \cosh \left(\alpha / 2 - \alpha R_{\alpha}\right)} \notag + \frac{1}{1 - e^{\alpha \left(1 - R_{\alpha}\right)}}.
\end{align}

where $n$ is the number of active compounds; $N$ is the total number of compounds; $R_{\alpha} = n/N$ is the ratio of the number of active compounds to the total number of compounds; $r_i$ is the ranking position of the $i^{th}$ active compound according to the scoring ranking.

Enrichment Factor (EF) is another metric to evaluate model performance, which calculates the fold increase in the proportion of active compounds among the top n\% of predicted compounds compared to the proportion of active compounds in the entire dataset. A higher EF value indicates better performance of the model in predicting active compounds.

$$
\text{EF}=\frac{\sum_{i=1}^n \delta_i}{\chi n} \quad \text { where } \delta_i= \begin{cases}1, & r_i \leq \chi N \\ 0, & r_{\mathrm{i}}>\chi N\end{cases}
$$
$\chi$ is the fraction of the ordered list that is considered and goes from 0 to 1. 

\subsection{Enzyme-Reaction Retrieval}

Here we introduce the metrics we use to evaluate the efficiency of the enzyme-reaction retrieval task. We omit the Top 2/3/4 and Top 2/3/4-N metrics from ReactZyme in Table \ref{tab:time.split}, \ref{tab:enzyme.split} and \ref{tab:reaction.split} to achieve better formatting,

Top-k accuracy measures whether the true positive label appears within the top \(k\) ranked predictions. The formula is as follows:

\[
\text{Top-k} = \frac{1}{N} \sum_{i=1}^{N} \frac{\sum_{j=1}^{k} \mathbb{I}(\text{Rank}_{ij} \leq k)}{k}
\]

where \(N\) is the total number of samples, \(\text{Rank}_{ij}\) is the rank of the \(j\)-th true label in the \(i\)-th sample, \(\mathbb{I}(\cdot)\) is the indicator function, which equals 1 if the condition is true and 0 otherwise, and \(k\) is the top-k cutoff threshold.

Top-k-N accuracy measures the fraction of sequences where at least one true label is present in the top \(k\) predictions. The formula is as follows:

\[
\text{Top-k-N} = \frac{1}{N} \sum_{i=1}^{N} \mathbb{I} \left( \sum_{j=1}^{k} \mathbb{I}(\text{Rank}_{ij} \leq k) > 0 \right)
\]

Mean Rank calculates the average rank of all true labels across all samples, capturing the overall ranking performance. The formula is as follows:

\[
\text{Mean Rank} = \frac{1}{N} \sum_{i=1}^{N} \frac{\sum_{j=1}^{M_i} \text{Rank}_{ij}}{M_i}
\]

where \(M_i\) is the number of true labels for the \(i\)-th sample.

Mean Reciprocal Rank (MRR) measures the average reciprocal rank of the first correct prediction for each sample, rewarding early correct predictions. The formula is as follows:

\[
\text{MRR} = \frac{1}{N} \sum_{i=1}^{N} \frac{1}{\min_j(\text{Rank}_{ij})}
\]

where \(\min_j(\text{Rank}_{ij})\) is The rank of the first correct prediction for the \(i\)-th sample, where \(j\) iterates over all true labels for that sample.

\section{Proof for FGW-CLIP}
\label{app:FGW-CLIP-Framework}

First, we establish a lemma for the loss \ref{eq lre} induced by the inverse optimal transport problem.

\begin{lemma}
\label{IOT-CL}
The loss in Equation \ref{eq lre} can be derived from the following optimization problem:
\begin{equation}
\label{kl-ot}
\begin{split}  
& \min_{\theta} \quad KL(\hat{P}||P^{\theta}) \\
& \text{subject to} \quad   
   P^{\theta} = \arg\min_{P \in U(a)} \left( \langle C^{\theta}, P \rangle - \tau H(P) \right)  
\end{split}  
\end{equation}
 where \(C^{\theta} \in R^{N \times N}, C^{\theta}(i, j) = c - s_{ij}(\theta)\) and \(\hat{P}(i, j) = \frac{f_{ij}}{N}\), where \(f_{ij}\) are label which equals to 1 when \(x_{i}\) and \(x_{j}\) are related else 0. \(I_{i}\) denotes the index associated with \(x_{i}\). \( U(\mathbf{a}) = \{\Gamma \in R_{+}^{N \times N} | \Gamma \mathbf{1}_N = \mathbf{a}\} \) . Here \(\mathbf{a}\) denotes a vector whose elements are the sums of labels, specifically defined as \(\mathbf{a}(i) = \sum_{j=1}^{N}f_{ij} \).
\end{lemma}

We introduce the Lagrangian of equation \ref{IOT-CL} as follows:

\begin{equation}
L(P, d) = ( \langle C^{\theta}, P \rangle - \tau H(P) - \sum_{i=1}^{N} d_{i}(\sum_{j=1}^{N}(P_{ij} - a_{i}))
\end{equation}

The KKT conditions can be obtained as follows:
\begin{equation}
\frac{\partial L(P, d)}{\partial P_{ij}} = C_{ij}^{\theta} + \tau log P_{ij} - d_{i} = 0
\end{equation}
Given that \(\sum_{j=1}^{N}P_{ij} = a_{i}\), we can derive the following expression:
\begin{equation}
    P_{ij} = \frac{a_{i} e^{-C_{ij}^{\theta}/\tau}}{\sum_{j=1}^{N}e^{-C_{ij}^{\theta}/\tau}}
\end{equation}

By solving the optimization problem \ref{kl-ot} according to definition, we can obtain the following results:
\begin{equation}
L_{iot} = -\frac{1}{N}\sum_{i=1}^{N}a_{i}log(\frac{\sum_{j \in I_{i}}e^{-C_{ij}^{\theta}/\tau}}{\sum_{j=1}^{N}e^{-C_{ij}^{\theta}/\tau}}) + Constant
\end{equation}
To simplify the problem, we disregard \(a_{i}\) and constant in \(L_{iot}\)  in practical applications, resulting in the following expression:
\begin{equation}
L_{iot} = -\sum_{i=1}^{N}log(\frac{\sum_{j \in I_{i}}e^{-C_{ij}^{\theta}/\tau}}{\sum_{j=1}^{N}e^{-C_{ij}^{\theta}/\tau}})
\end{equation}

\subsection{Proof for Proposition \ref{FGW-CLIP-Framework}}
\label{FGW-F}

According to Lemma \ref{IOT-CL}, we can transform the original optimization problem into the following problem:

\begin{equation}  
\label{FGW-CLIP_framework_induced_1}  
\begin{aligned}  
& \min_{\theta, \psi_{1}, \psi_{2}} \Big\{ -(1-\alpha) \sum_{i=1}^{N} \sum_{j \in I_{i}} \log \left( \frac{e^{-C_{ij}^{\theta}/\tau}}{\sum_{j=1}^{N} e^{-C_{ij}^{\theta}/\tau}} \right) + \alpha GW(\Gamma^{\psi_{1}}_d, \Gamma^{\psi_{2}}_d, \Gamma^{\theta}) \\
& \quad - \lambda_{1} \sum_{i=1}^{N} \sum_{j \in I_{i}} \log \left( \frac{e^{-C_{ij}^{\psi_{1}}/\tau}}{\sum_{j=1}^{N} e^{-C_{ij}^{\psi_{1}}/\tau}} \right) - \lambda_{2} \sum_{i=1}^{N} \sum_{j \in I_{i}} \log \left( \frac{e^{-C_{ij}^{\psi_{2}}/\tau}}{\sum_{j=1}^{N} e^{-C_{ij}^{\psi_{2}}/\tau}} \right) \quad - \lambda_{ce} CE(y_{\psi_{1}}, f_{\psi_{1}}(X_{1})) \Big\}  
\end{aligned}  
\end{equation}

For the GW term, we can simplify it according to the definition as follows:

\begin{equation}
\begin{aligned}
    GW(\Gamma^{\psi_{1}}_d, \Gamma^{\psi_{2}}_d, \Gamma^{\theta}) = \, \left( \Gamma^{\psi_{1}}_d \circ \Gamma^{\psi_{1}}_d a^{\psi_{1}} \right)^{\top} a^{\psi_{1}} 
     + \left( \Gamma^{\psi_{2}}_d \circ \Gamma^{\psi_{2}}_d a^{\psi_{2}} \right)^{\top} a^{\psi_{2}} 
     - 2 \, \text{tr} \left( \left( \Gamma^{\theta} \right)^{\top} \Gamma_{d}^{\psi_{1}} \Gamma^{\theta} \Gamma_{d}^{\psi_{2}} \right)
\end{aligned}
\end{equation}

which \(\circ\) Hadamard product.
Disregarding the constant terms, we can simplify the optimization objective as follows:
\begin{equation}
    GW(\Gamma^{\psi_{1}}_d, \Gamma^{\psi_{2}}_d, \Gamma^{\theta}) =  - 2 tr\left((\Gamma^{\theta})^{\top} \Gamma_{d}^{\psi_{1}} \Gamma^{\theta} \Gamma_{d}^{\psi_{2}}\right)
\end{equation}

Considering the symmetric positions of 
 \(i\) and  \(j\), a classic technique is to transform the original optimization problem into the following form:

\begin{equation}  
\label{FGW-CLIP_framework_induced_2}  
\begin{aligned}  
& \min_{\theta, \psi_{1}, \psi_{2}} \Big\{ -\frac{1-\alpha}{2} \Big( \sum_{i=1}^{N} \sum_{j \in I_{i}} \log \left( \frac{e^{-C_{ij}^{\theta}/\tau}}{\sum_{j=1}^{N} e^{-C_{ij}^{\theta}/\tau}} \right) \quad + \sum_{j=1}^{N} \sum_{i \in J_{j}} \log \left( \frac{e^{-C_{ij}^{\theta}/\tau}}{\sum_{i=1}^{N} e^{-C_{ij}^{\theta}/\tau}} \right) \Big) - 2 \alpha \, \text{tr} \left( (\Gamma^{\theta})^{\top} \Gamma_{d}^{\psi_{1}} \Gamma^{\theta} \Gamma_{d}^{\psi_{2}} \right) \\
& \quad - \lambda_{1} \sum_{i=1}^{N} \sum_{j \in I_{i}} \log \left( \frac{e^{-C_{ij}^{\psi_{1}}/\tau}}{\sum_{j=1}^{N} e^{-C_{ij}^{\psi_{1}}/\tau}} \right) - \lambda_{2} \sum_{i=1}^{N} \sum_{j \in I_{i}} \log \left( \frac{e^{-C_{ij}^{\psi_{2}}/\tau}}{\sum_{j=1}^{N} e^{-C_{ij}^{\psi_{2}}/\tau}} \right) \quad - \lambda_{ce} \, \text{CE} \left( y_{\psi_{1}}, f_{\psi_{1}}(X_{1}) \right) \Big\}
\end{aligned}  
\end{equation}

The above expression is consistent with the form of the overall loss obtained by FGW-CLIP.
Given that \( i \in I_{i}\), \( j \in J_{j}\) and \(C^{\theta}(i,j) = c -x_{\psi_{1},i}x_{\psi_{2},j}^{T}\), by reorganizing equation \ref{FGW-CLIP_framework_induced_2}, it can be observed that:

\begin{equation}
\begin{aligned}  
& \min_{\theta, \psi_{1}, \psi_{2}} \Big\{ -\frac{1-\alpha}{2} \Big( \sum_{i=1}^{N} \sum_{j \in I_{i}} \frac{x_{\psi_{1},i} x_{\psi_{2},j}^{T} - c}{\tau} + \sum_{j=1}^{N} \sum_{i \in J_{j}} \frac{x_{\psi_{2},i} x_{\psi_{1},j}^{T} - c}{\tau} \Big) \quad + \alpha \, GW(\Gamma^{\psi_{1}}_d, \Gamma^{\psi_{2}}_d, \Gamma^{\theta}) \\
& + \frac{1-\alpha}{2} \Big( \sum_{i=1}^{N} \sum_{j \in I_{i}} \log \left( \sum_{j=1}^{N} e^{-C_{ij}^{\theta}/\tau} \right) \quad + \sum_{j=1}^{N} \sum_{i \in J_{j}} \log \left( \sum_{i=1}^{N} e^{-C_{ij}^{\theta}/\tau} \right) \Big) \\
& \quad - \lambda_{1} \sum_{i=1}^{N} \sum_{j \in I_{i}} \log \left( \frac{e^{-C_{ij}^{\psi_{1}}/\tau}}{\sum_{j=1}^{N} e^{-C_{ij}^{\psi_{1}}/\tau}} \right)
- \lambda_{2} \sum_{i=1}^{N} \sum_{j \in I_{i}} \log \left( \frac{e^{-C_{ij}^{\psi_{2}}/\tau}}{\sum_{j=1}^{N} e^{-C_{ij}^{\psi_{2}}/\tau}} \right) \quad - \lambda_{ce} \, \text{CE} \left( y_{\psi_{1}}, f_{\psi_{1}}(X_{1}) \right) \Big\}
\end{aligned} 
\end{equation}

Since \(x_{\psi_{1}}, x_{\psi_{2}}\) are \(L_{2}\) normalized, disregarding constants, we can derive that:

\begin{equation}
\begin{aligned}  
& \min_{\theta, \psi_{1}, \psi_{2}} \Big\{ \frac{1-\alpha}{2} \Big( \sum_{i=1}^{N} \sum_{j \in I_{i}} \frac{|x_{\psi_{1},i} - x_{\psi_{2},j}|^{2}}{\tau^{2}} + \sum_{j=1}^{N} \sum_{i \in J_{j}} \frac{|x_{\psi_{2},i} - x_{\psi_{1},j}|^{2}}{\tau^{2}} \Big) \quad + \alpha GW(\Gamma^{\psi_{1}}_d, \Gamma^{\psi_{2}}_d, \Gamma^{\theta}) \\
& + \frac{1-\alpha}{2} \Big( \sum_{i=1}^{N} \sum_{j \in I_{i}} \log \left( \sum_{j=1}^{N} e^{-C_{ij}^{\theta}/\tau} \right) \quad + \sum_{j=1}^{N} \sum_{i \in J_{j}} \log \left( \sum_{i=1}^{N} e^{-C_{ij}^{\theta}/\tau} \right) \Big) \\
& \quad - \lambda_{1} \sum_{i=1}^{N} \sum_{j \in I_{i}} \log \left( \frac{e^{-C_{ij}^{\psi_{1}}/\tau}}{\sum_{j=1}^{N} e^{-C_{ij}^{\psi_{1}}/\tau}} \right) - \lambda_{2} \sum_{i=1}^{N} \sum_{j \in I_{i}} \log \left( \frac{e^{-C_{ij}^{\psi_{2}}/\tau}}{\sum_{j=1}^{N} e^{-C_{ij}^{\psi_{2}}/\tau}} \right) \quad - \lambda_{ce} \, \text{CE} \left( y_{\psi_{1}}, f_{\psi_{1}}(X_{1}) \right) \Big\}
\end{aligned} 
\end{equation}

From the equation, we can deduce that \(L_FGW\) is the optimization of a specific fused Gromov-Wasserstein distance under regularization conditions.

\section{Experiment Details}
For the training of FGW-CLIP on EnzymeMap, we use the Adam optimizer with $\beta_1 = 0.9$, $\beta_2 = 0.99$, and $\epsilon = 10^{-8}$, along with a weight decay of $1 \times 10^{-4}$. The initial learning rate is 0.001, and we apply a polynomial decay scheduler with a warmup ratio of 0.06. The batch size is 32, distributed across 4 NVIDIA GeForce RTX 4090 24G GPUs. The dropout rate is set to 0.1. The temperature parameter $\tau$ used in all contrastive losses is 0.05. The weight $\lambda$ for the EC prediction loss $L_{\text{EC}}$ is set to 0.1. For the reaction branch, we adopt Uni-Mol as the molecular encoder, with the same hyperparameters as in its original paper. The molecular readout function is set to sum. For the enzyme branch, we use the pretrained ESM2 protein language model as the encoder. ESM2 is frozen during training, and we append a linear projection head to map its output to the shared representation space. The training epoch is 100, and the last checkpoint is selected.

For training on ReactZyme, the number of epochs is set to 50, and the best model checkpoint is selected based on validation loss. All experiments on ReactZyme are conducted using 4 NVIDIA A100-SXM4-80GB GPUs. The remaining parameters are the same as those used during EnzymeMap training.

\subsection{Computational Complexity of GW Distance}
\label{app:GW}

The computation of the GW distance in Eq.~\ref{eq GW} involves a quadruple summation over index pairs \((i, j, i', j')\), leading to a theoretical time complexity of \(\mathcal{O}(N^4)\), where \(N\) is the batch size. Despite the theoretical cost, this computation is applied only within mini-batches, rather than over the entire dataset. As a result, the actual runtime is independent of the total dataset size.

To empirically evaluate this, we benchmarked our method with and without GW regularization on the EnzymeMap dataset. Results show:
\begin{itemize}
    \item The per-step runtime increases by only 0.0037 seconds, a relative increase of 3.3\%;
    \item GPU memory consumption remains virtually unchanged on a single GPU.
\end{itemize}

These findings indicate that, although GW has high theoretical complexity, the practical overhead is minimal in our mini-batch training scheme.

\section{The results on the enzyme-similarity-based split and reaction-similarity-based split of ReactZyme}
\label{app:react}

\begin{table*}[htbp!]

\centering
\caption{Retrieval Performance on ReactZyme of the enzyme-similarity-based split. The best results are \textbf{bolded}. Except for the mean rank, the higher the metric, the better. The baselines marked with * indicate that they use FANN to enhance residue-level representations.}
\label{tab:enzyme.split}

\subfloat[Given the enzyme, retrieve positive reactions.]{
\resizebox{\columnwidth}{!}{%

\begin{tabular}{l|cccc|cccc|c|c}
\toprule
Enzyme/enzyme-reaction & Top1 & Top5 & Top10 & Top20 & Top1-N & Top5-N & Top10-N & Top20-N & Mean Rank & MRR \\
\midrule
Data(Ground-truth) & 1.0000 & 1.0000 & 1.0000 & 1.0000 & 1.0000 & 0.2001 & 0.1001 & 0.0500 & 1.0003 & 0.9999 \\
\midrule
MAT-2D + ESM & 0.5987 & 0.8759 & 0.9328 & 0.9572 & 0.5987 & 0.1774 & 0.0939 & 0.0485 & 5.3021 & 0.7280 \\
MAT-2D + SaProt & 0.6691 & 0.8893 & 0.9358 & 0.9553 & 0.6691 & 0.1801 & 0.0942 & 0.0484 & 5.4356 & 0.7733 \\
UniMol-2D + ESM & 0.6077 & 0.8759 & 0.9338 & 0.9533 & 0.6077 & 0.1774 & 0.0940 & 0.0483 & 7.0311 & 0.7349 \\
UniMol-2D + SaProt & 0.5717 & 0.8357 & 0.8891 & 0.9474 & 0.5717 & 0.1693 & 0.0895 & 0.0480 & 15.2646 & 0.6912 \\
UniMol-2D + ESM* & 0.6256 & 0.8749 & 0.9348 & 0.9493 & 0.6256 & 0.1772 & 0.0941 & 0.0481 & 7.0024 & 0.7491 \\
UniMol-2D + SaProt* & 0.6038 & 0.8695 & 0.9338 & 0.9375 & 0.6038 & 0.1761 & 0.0940 & 0.0475 & 7.0746 & 0.7346 \\
\midrule
MAT-3D + ESM & 0.4544 & 0.6408 & 0.8573 & 0.9118 & 0.4544 & 0.1300 & 0.0863 & 0.0462 & 30.8473 & 0.5093 \\
MAT-3D + SaProt & 0.5539 & 0.6904 & 0.8821 & 0.9296 & 0.5539 & 0.1400 & 0.0888 & 0.0471 & 15.3962 & 0.6735 \\
UniMol-3D + ESM & 0.7267 & 0.9062 & 0.9487 & 0.9632 & 0.7267 & 0.1835 & 0.0955 & 0.0488 & 4.5799 & 0.8112 \\
UniMol-3D + SaProt & 0.5998 & 0.8665 & 0.9229 & 0.9454 & 0.5998 & 0.1755 & 0.0929 & 0.0479 & 7.4701 & 0.7226 \\
UniMol-3D + ESM* & 0.7111 & 0.9017 & 0.9547 & 0.9592 & 0.7111 & 0.1826 & 0.0961 & 0.0486 & 4.8395 & 0.8023 \\
UniMol-3D + SaProt* & 0.6328 & 0.8853 & 0.9348 & 0.9513 & 0.6328 & 0.1793 & 0.0941 & 0.0482 & 6.9597 & 0.7457 \\
\midrule
CLIPZyme (MAT-2D + ESM)& 0.5489 & 0.7768 & 0.9290 & 0.9460 & 0.5489 & 0.1554 & 0.0929 & 0.0473 & 8.3524 & 0.6971 \\
CLIPZyme (UniMol-3D + ESM) & 0.7547 & 0.9478 & 0.9679 & 0.9780 & 0.7547 & 0.1896 & 0.0968 & 0.0489 & 3.9820 & 0.8546 \\
\midrule
FGW-CLIP, EC Mode & \textbf{0.7875} & 0.9742 & \textbf{0.9891} & \textbf{0.9953} & \textbf{0.7875} & 0.1949 & \textbf{0.0990} & \textbf{0.0498} & 2.4126 & \textbf{0.8683} \\
FGW-CLIP, EC Max & 0.7642 & \textbf{0.9746} & 0.9875 & 0.9953 & 0.7642 & \textbf{0.1950} & 0.0988 & 0.0498 & \textbf{2.3837} & 0.8581 \\
\bottomrule

    \end{tabular}%
}}

\subfloat[Given the reaction, retrieve positive enzymes.]{

\resizebox{\columnwidth}{!}{%
    \begin{tabular}{l|cccc|cccc|c|c}
\toprule
Enzyme/reaction-enzyme & Top1 & Top5 & Top10 & Top20 & Top1-N & Top5-N & Top10-N & Top20-N & Mean Rank & MRR \\
\midrule
Ground-truth & 1.0000 & 1.0000 & 1.0000 & 1.0000 & 1.0000 & 0.4833 & 0.3370 & 0.2263 & 3.2778 & 0.7321 \\
\midrule
MAT-2D + ESM & 0.3624 & 0.6091 & 0.7225 & 0.7986 & 0.3624 & 0.2961 & 0.2444 & 0.1820 & 22.5053 & 0.2586 \\
MAT-2D + SaProt & 0.3999 & 0.6455 & 0.7583 & 0.8390 & 0.3999 & 0.3138 & 0.2565 & 0.1912 & 23.5890 & 0.2883 \\
UniMol-2D + ESM & 0.3435 & 0.5701 & 0.7007 & 0.7530 & 0.3435 & 0.2771 & 0.2370 & 0.1716 & 25.4892 & 0.2512 \\
UniMol-2D + SaProt & 0.3049 & 0.5273 & 0.6347 & 0.6872 & 0.3049 & 0.2563 & 0.2147 & 0.1566 & 30.5631 & 0.2245 \\
UniMol-2D + ESM* & 0.3584 & 0.5892 & 0.7338 & 0.7543 & 0.3584 & 0.2864 & 0.2482 & 0.1719 & 25.0362 & 0.2674 \\
UniMol-2D + SaProt* & 0.3534 & 0.5713 & 0.7051 & 0.7640 & 0.3534 & 0.2777 & 0.2385 & 0.1741 & 25.1678 & 0.2635 \\
\midrule
MAT-3D + ESM & 0.3827 & 0.6373 & 0.7089 & 0.8048 & 0.3827 & 0.3098 & 0.2398 & 0.1834 & 26.4117 & 0.2841 \\
MAT-3D + SaProt & 0.3751 & 0.5493 & 0.6572 & 0.7394 & 0.3751 & 0.2670 & 0.2223 & 0.1685 & 24.5678 & 0.2763 \\
UniMol-3D + ESM & 0.4088 & 0.6892 & 0.7953 & 0.8666 & 0.4088 & 0.3350 & 0.2690 & 0.1975 & 24.2505 & 0.2930 \\
UniMol-3D + SaProt & 0.3477 & 0.5458 & 0.6980 & 0.7762 & 0.3477 & 0.2653 & 0.2361 & 0.1769 & 34.9487 & 0.2562 \\
UniMol-3D + ESM* & 0.3928 & 0.6612 & 0.7628 & 0.8324 & 0.3928 & 0.3214 & 0.2580 & 0.1897 & 23.8241 & 0.2837 \\
UniMol-3D + SaProt* & 0.3655 & 0.6161 & 0.7376 & 0.7552 & 0.3655 & 0.2995 & 0.2495 & 0.1721 & 22.8901 & 0.2633 \\
\midrule
CLIPZyme (MAT-2D + ESM) & 0.3337 & 0.6077 & 0.6514 & 0.7687 & 0.3337 & 0.2844 & 0.2235 & 0.1811 & 30.4196 & 0.2038 \\
CLIPZyme (UniMol-3D + ESM) & 0.3570 & 0.6371 & 0.7552 & 0.8431 & 0.3570 & 0.2885 & 0.2577 & 0.1834 & 25.5786 & 0.2828 \\
\midrule
FGW-CLIP, EC Mode & 0.6940 & 0.8893 & 0.9319 & 0.9701 & 0.6940 & 0.4327 & 0.3176 & 0.2213 & 11.3119 & 0.5253 \\
FGW-CLIP, EC Max & \textbf{0.6978} & \textbf{0.8931} & \textbf{0.9389} & \textbf{0.9739} & \textbf{0.6978} & \textbf{0.4338} & \textbf{0.3200} & \textbf{0.2216} & \textbf{10.8511} & \textbf{0.5300} \\

\bottomrule
    \end{tabular}%
}}

\end{table*}

\begin{table*}[htbp!]

\centering
\caption{Retrieval Performance on ReactZyme of the reaction-similarity-based split. The best results are \textbf{bolded}. Except for the mean rank, the higher the metric, the better. The baselines marked with * indicate that they use FANN to enhance residue-level representations.}
\label{tab:reaction.split}

\subfloat[Given the enzyme, retrieve positive reactions.]{
\resizebox{\columnwidth}{!}{%

\begin{tabular}{l|cccc|cccc|c|c}
\toprule
Reaction/enzyme-reaction & Top1 & Top5 & Top10 & Top20 & Top1-N & Top5-N & Top10-N & Top20-N & Mean Rank & MRR \\
\midrule
Data(Ground-truth) & 1.0000 & 1.0000 & 1.0000 & 1.0000 & 1.0000 & 0.2000 & 0.1000 & 0.0500 & 1.0000 & 1.0000 \\
\midrule
MAT-2D + ESM & 0.0914 & 0.2968 & 0.4373 & 0.5908 & 0.0914 & 0.0596 & 0.0438 & 0.0296 & 39.9146 & 0.2005 \\
MAT-2D + SaProt & 0.0963 & 0.3018 & 0.4123 & 0.5070 & 0.0963 & 0.0606 & 0.0413 & 0.0254 & 72.0597 & 0.1936 \\
UniMol-2D + ESM & 0.0949 & 0.2694 & 0.4363 & 0.4232 & 0.0949 & 0.0541 & 0.0437 & 0.0212 & 65.2719 & 0.1865 \\
UniMol-2D + SaProt & 0.0944 & 0.2754 & 0.4143 & 0.4571 & 0.0944 & 0.0553 & 0.0415 & 0.0229 & 59.7940 & 0.1956 \\
UniMol-2D + ESM & 0.0929 & 0.2610 & 0.4313 & 0.4271 & 0.0929 & 0.0524 & 0.0432 & 0.0214 & 72.7932 & 0.1810 \\
UniMol-2D + SaProt & 0.0926 & 0.2699 & 0.4343 & 0.5309 & 0.0926 & 0.0542 & 0.0435 & 0.0266 & 89.8456 & 0.1857 \\
\midrule
MAT-3D + ESM & 0.0930 & 0.2595 & 0.4203 & 0.4431 & 0.0930 & 0.0521 & 0.0421 & 0.0222 & 81.3234 & 0.1893 \\
MAT-3D + SaProt & 0.0915 & 0.2565 & 0.4293 & 0.5269 & 0.0915 & 0.0515 & 0.0430 & 0.0264 & 94.9242 & 0.1804 \\
UniMol-3D + ESM & 0.0912 & 0.2580 & 0.4213 & 0.4571 & 0.0912 & 0.0518 & 0.0422 & 0.0229 & 92.2778 & 0.1856 \\
UniMol-3D + SaProt & 0.1085 & 0.2699 & 0.4034 & 0.5429 & 0.1085 & 0.0542 & 0.0404 & 0.0272 & 42.3597 & 0.1988 \\
UniMol-3D + ESM & 0.1104 & 0.3023 & 0.4573 & 0.5669 & 0.1104 & 0.0607 & 0.0458 & 0.0284 & 38.9685 & 0.2011 \\
UniMol-3D + SaProt & 0.0962 & 0.2545 & 0.4024 & 0.5289 & 0.0962 & 0.0511 & 0.0403 & 0.0265 & 50.9663 & 0.1972 \\
\midrule
CLIPZyme (MAT-2D + ESM) & 0.1235 & 0.3064 & 0.5719 & 0.6000 & 0.1235 & 0.0613 & 0.0572 & 0.0300 & \textbf{35.6457} & 0.2201 \\
CLIPZyme (UniMol-3D + ESM) & 0.1305 & 0.3420 & 0.5320 & 0.6220 & 0.1305 & 0.0684 & 0.0532 & 0.0311 & 48.4672 & 0.1937 \\
\midrule
FGW-CLIP, EC Mode  & 0.2486 & 0.5088 & 0.5716 & 0.6774 & 0.2486 & 0.1018 & 0.0572 & 0.0339 & 49.5410 & 0.3722 \\
FGW-CLIP, EC Max & \textbf{0.2561} & \textbf{0.5117} & \textbf{0.5799} & \textbf{0.6919} & \textbf{0.2561} & \textbf{0.1024} & \textbf{0.0580} & \textbf{0.0346} & 41.8125 & \textbf{0.3804} \\

\bottomrule
    \end{tabular}%
}}

\subfloat[Given the reaction, retrieve positive enzymes.]{
\resizebox{\columnwidth}{!}{%

\begin{tabular}{l|cccc|cccc|c|c}
\toprule
Reaction/reaction-enzyme & Top1 & Top5 & Top10 & Top20 & Top1-N & Top5-N & Top10-N & Top20-N & Mean Rank & MRR \\
\midrule
Ground-truth & 1.0000 & 1.0000 & 1.0000 & 1.0000 & 1.0000 & 0.5389 & 0.3870 & 0.2711 & 19.5272 & 0.6715 \\
\midrule
MAT-2D + ESM & 0.1347 & 0.2000 & 0.2326 & 0.2753 & 0.1347 & 0.1083 & 0.0902 & 0.0749 & 529.4258 & 0.1341 \\
MAT-2D + SaProt & 0.0933 & 0.1627 & 0.2213 & 0.2561 & 0.0933 & 0.0881 & 0.0858 & 0.0697 & 504.8481 & 0.1076 \\
UniMol-2D + ESM & 0.0931 & 0.1321 & 0.1769 & 0.1863 & 0.0931 & 0.0715 & 0.0686 & 0.0507 & 550.0562 & 0.0946 \\
UniMol-2D + SaProt & 0.0910 & 0.1380 & 0.1818 & 0.2345 & 0.0910 & 0.0747 & 0.0705 & 0.0638 & 567.8300 & 0.0989 \\
UniMol-2D + ESM & 0.1033 & 0.1502 & 0.2076 & 0.2547 & 0.1033 & 0.0813 & 0.0805 & 0.0693 & 590.4462 & 0.0928 \\
UniMol-2D + SaProt & 0.0905 & 0.1339 & 0.1813 & 0.2407 & 0.0905 & 0.0725 & 0.0703 & 0.0655 & 549.8296 & 0.0961 \\
\midrule
MAT-3D + ESM & 0.1269 & 0.1962 & 0.2251 & 0.2712 & 0.1269 & 0.1062 & 0.0873 & 0.0738 & 532.6187 & 0.1184 \\
MAT-3D + SaProt & 0.0909 & 0.1407 & 0.1849 & 0.2528 & 0.0909 & 0.0762 & 0.0717 & 0.0688 & 539.1481 & 0.1044 \\
UniMol-3D + ESM & 0.0924 & 0.1332 & 0.1790 & 0.2172 & 0.0924 & 0.0721 & 0.0694 & 0.0591 & 548.3340 & 0.0943 \\
UniMol-3D + SaProt & 0.0933 & 0.1703 & 0.2130 & 0.2613 & 0.0933 & 0.0922 & 0.0826 & 0.0711 & \textbf{493.1189} & 0.0962 \\
UniMol-3D + ESM & 0.1244 & 0.2058 & 0.2440 & 0.2848 & 0.1244 & 0.1114 & 0.0946 & 0.0775 & 559.1225 & 0.1129 \\
UniMol-3D + SaProt & 0.0917 & 0.1418 & 0.1847 & 0.2234 & 0.0917 & 0.0768 & 0.0716 & 0.0608 & 552.4546 & 0.1051 \\
\midrule
CLIPZyme (MAT-2D + ESM) & 0.1457 &0.2173 & 0.2456 & 0.2893 & 0.1457 & 0.1135 & 0.1001 & 0.0783 & 501.2071 & 0.1521 \\
CLIPZyme (UniMol-3D + ESM) & 0.1298 & 0.1993 & 0.2215 & 0.2544 & 0.1298 & 0.0866 & 0.0830 & 0.0741 & 526.4793 & 0.1245 \\
\midrule
FGW-CLIP, EC Mode &  \textbf{0.3886} & 0.5907 & \textbf{0.6736} & \textbf{0.7228} & \textbf{0.3886} & 0.2705 & 0.2091 & 0.1492 & 567.1841 & \textbf{0.3181} \\
FGW-CLIP, EC Max & 0.3782 & \textbf{0.5933} & 0.6658 & 0.7150 & 0.3782 & \textbf{0.2746} & \textbf{0.2163} & \textbf{0.1530} & 552.7454 & 0.3113 \\
\bottomrule
    \end{tabular}
}}
\end{table*}

We provide detailed results for the enzyme and reaction retrieval experiments on the enzyme similarity-based and reaction similarity-based splits of ReactZyme in Tables \ref{tab:enzyme.split} and \ref{tab:reaction.split}, respectively. The experiments follow the same setup described for the time-based split in the main text, with FGW-CLIP compared against the baselines. The results for CLIPZyme in Table \ref{tab:time.split} of the main text and Tables \ref{tab:enzyme.split} and \ref{tab:reaction.split} here are sourced from Tables 15, 16, and 17 of the ReactZyme paper.

For the enzyme similarity split (Table \ref{tab:enzyme.split}), FGW-CLIP demonstrates superior performance across both retrieval tasks. In the reaction similarity split (Table \ref{tab:reaction.split}), FGW-CLIP achieves SOTA results on all metrics except Mean Rank. The lower Mean Rank may be attributed to the limited number of unique reactions (7,726) compared to enzymes (178,327) in ReactZyme, leading to some reactions being ranked lower during retrieval. Despite this, the impact on Top-k metrics is minimal, showcasing the robustness of FGW-CLIP in retrieval tasks with no reaction overlap between training and test sets.

Experiments in Tables \ref{tab:enzyme.split} and \ref{tab:reaction.split} validate the findings presented in the main text, further emphasizing FGW-CLIP’s effectiveness and its consistent outperformance of baselines, including those using the same encoders.

\subsection{Limitation}
Currently, the framework scenario is only limited to the case of enzyme screening. In the future, it will be applied to more extensive enzyme prediction and design tasks.

\section{Potential societal impacts}
This work contributes to enzyme screening and functional annotation, with potential positive societal impacts in biocatalysis, drug discovery, and sustainable chemistry. By improving the identification of useful enzymes, FGW-CLIP can help accelerate biological research and enable environmentally friendly industrial applications.

As with any machine learning model in the biochemical domain, there is a potential risk of misuse in synthetic biology without proper oversight. While our work is intended for scientific research, we recommend that any downstream application adhere to ethical standards and regulatory frameworks to ensure responsible use.